\documentclass[reprint,amsmath,amssymb,aps,prb,superscriptaddress,longbibliography]{revtex4-1}
\usepackage{array}
\usepackage{bm}
\usepackage{color}
\usepackage{float}
\usepackage{graphicx}
\usepackage[colorlinks,citecolor=blue]{hyperref}
\usepackage{natbib}
\usepackage{newtxtext}
\usepackage{newtxmath}
\usepackage[caption=false]{subfig}
\usepackage{mathrsfs}

\usepackage{amsmath,amssymb}
\usepackage{verbatim}
\usepackage{amsfonts}
\usepackage{dcolumn}
\usepackage{bm}
\usepackage{color}

\usepackage{float}
\usepackage{pifont}

\newcommand{\beq}{\begin{eqnarray} }
\newcommand{\eeq}{\end{eqnarray} }
\newcommand{\Beq}{\begin{eqnarray*} }
\newcommand{\Eeq}{\end{eqnarray*} }
\newcommand{\Bmat}{\left(\begin{matrix}}
\newcommand{\Emat}{\end{matrix}\right)}
\newcommand{\up}{\uparrow}
\newcommand{\dn}{\downarrow}

\begin{document}

\title{Multinode quantum spin liquids on the honeycomb lattice }
\author{Jiucai Wang}
\affiliation{Department of Physics, Renmin University of China, Beijing 100872, China}

\author{Qirong Zhao}
\affiliation{Department of Physics, Renmin University of China, Beijing 100872, China}

\author{Xiaoqun Wang}
\email{xiaoqunwang@sjtu.edu.cn}
\affiliation{Key Laboratory of Artificial Structures and Quantum Control of MOE, Shenyang National Laboratory for Materials Science, Shenyang 110016, China}
\affiliation{School of Physics and Astronomy, Tsung-Dao Lee Institute, Shanghai Jiao Tong University, Shanghai 200240, China}

\author{Zheng-Xin Liu}
\email{liuzxphys@ruc.edu.cn}
\affiliation{Department of Physics, Renmin University of China, Beijing 100872, China}
\affiliation{School of Physics and Astronomy, Tsung-Dao Lee Institute, Shanghai Jiao Tong University, Shanghai 200240, China}

\date{\today}

\begin{abstract}
Recently it was realized that the zigzag magnetic order in Kitaev materials can be stabilized by small negative off-diagonal interactions called the $\Gamma'$ terms. To fully understand the effect of the $\Gamma'$ interactions, we investigate the quantum $K$-$\Gamma$-$\Gamma'$ model on the honeycomb lattice using the variational Monte Carlo method.   Two multinode Z$_2$ quantum spin liquids (QSLs) are found at $\Gamma'>0$, one of which is the previously found proximate Kitaev spin liquid called the PKSL14 state which shares the same projective symmetry group (PSG) with the Kitaev spin liquid. A remarkable result is that a $\pi$-flux state with a distinct PSG appears at larger $\Gamma'$. The $\pi$-flux state is characterized by an enhanced periodic structure in the spinon dispersion in the original Brillouin zone (BZ), which is experimentally observable.
Interestingly, two PKSL8 states are competing with the $\pi$-flux state and one of them can be stabilized by six-spin ring-exchange interactions. The physical properties of these nodal QSLs are studied by applying magnetic fields and the results depend on the number of cones. Our study infers that there exist a family of zero-flux QSLs that contain $6n+2, n\in\mathbb Z$ Majorana cones and a family of $\pi$-flux QSLs containing $4(6n+2)$ cones in the original BZ. It provides guidelines for experimental realization of non-Kitaev QSLs in relevant materials.
\end{abstract}


\maketitle

\section{Introduction}
Quantum spin liquids (QSLs) are exotic phases of matter exhibiting no conventional long-range order down to the lowest temperatures \cite{Balents,ZhouYi}. However, it is challenging to construct lattice models to support spin-liquid ground states and then realize them in candidate materials. In 2006, Kitaev proposed a honeycomb lattice model which has an exactly solvable QSL ground state and a gapless or gapped excitation spectrum \cite{Kitaev}. In a general magnetic field, the gapless Kitaev spin liquid (KSL) can be turned into a gapped chiral spin liquid (CSL) that supports non-Abelian anyonic excitations. It has been proposed that spin-orbit entangled materials \cite{rjk,rcjk}, such as Na$_2$IrO$_3$ and $\alpha$-RuCl$_3$, contain Kitaev interactions ($S^\gamma_iS_j^\gamma$) and are candidates to realize the KSL. However, these materials manifest magnetic long-range order \cite{rsea,rjea,rcaoetal,ryea,rchoietal,Williams} at low temperatures, indicating that the non-Kitaev interactions such as the off-diagonal symmetric $\Gamma$ interactions ($S^\alpha_iS_j^\beta+S^\beta_iS_j^\alpha$) are not negligible \cite{lukas,Jinsheng}. Although many lattice models have been proposed as the effective interactions of the Kitaev materials \cite{npj}, none of them can explain all of the experimental data.
A third-neighbor Heisenberg interaction was proposed to interpret the zigzag order \cite{lukas, KJG_zigzag,You_zig,Thomale,Valenti}, but the question is that long-range interactions are usually too small to stabilize the order. Recently, another nearest-neighbor off-diagonal interaction ($S^\alpha_iS_j^\gamma + S^\gamma_iS_j^\alpha + S^\beta_iS_j^\gamma + S^\gamma_i S_j^\beta$) called the $\Gamma'$ term has attracted some attention \cite{Valenti,Rau}. Density matrix renormalization group and infinite tensor network studies have shown that a very small $\Gamma'<0$ can support a zigzag ordered ground state \cite{HYKee, tensor}. The physical origin of the $\Gamma'$ interactions may be owing to the trigonal distortion \cite{Valenti,Rau,hidden} of the Kitaev materials.  Since the parameters of the effective interactions in different materials are generally different, it is possible that in some compound the $\Gamma'$ interaction may switch its sign. The physical consequence of the interactions in such a parameter regime still needs to be revealed.

In the present work, the quantum $K$-$\Gamma$-$\Gamma'$ honeycomb model is studied using the variational Monte Carlo (VMC) method and the global phase diagram is obtained. Besides the well known KSL phase, we find two more QSLs for $\Gamma'>0$. One contains 14 Majorana cones in its excitation spectrum and shares the same projective symmetry group (PSG) \cite{igg,You_PSG} with the KSL, and is thus called the proximate-KSL14 (PKSL14) phase\cite{PKSL}. The other gapless QSL,  which contains 32 Majorana cones in the original Brillouin zone (BZ), has a distinct PSG since the spinons feel a uniform $\pi$ flux at each hexagon. The $\pi$-flux state has a sharp experimental character since there is an enhanced periodic structure in the spinon dispersion in the original BZ. In our VMC calculation, we find that two PKSL8 states are close in energy compared to the $\pi$-flux state, one of which can be stabilized by a small ring-exchange interaction. The physical responses to magnetic fields are found to be dependent on the number of cones. Our study reveals two families of QSLs respecting the $D_{3d}\times Z_2^T$ physical symmetry. The first one has zero-flux and contains $6n+2$ Majorana cones in its excitation spectrum, while the second one has a uniform $\pi$-flux and contains $4(6n+2)$ Majorana cones. This result is instructive for an experimental search of gapless QSLs in related materials.


The rest of the paper is organized as follows. In Sec.\ref{model}, we introduce an extended Kitaev model (the $K$-$\Gamma$-$\Gamma'$ model) on the honeycomb lattice and present the quantum phase diagram obtained from the VMC method.  Section \ref{Gutzwiller} is devoted to technical details of the VMC method, and is mainly focused on the way of constructing the Gutzwiller projected wave functions. Readers who are not interested in technical issues can skip Sec.\ref{Gutzwiller} and go to Sec.\ref{fields}, where we illustrate the physical response of the system to magnetic fields for the QSL phases as well as the magnetically ordered phases. The paper is concluded in Sec.\ref{conclusion}, before which we discuss the possible existence of two families of multinode QSL phases.

\section{The model and  the phase diagram}\label{model} 
We start with the extended Kitaev honeycomb model containing $K$, $\Gamma$ and $\Gamma'$ interactions,
\begin{eqnarray}\label{KGG}
H =  \sum_{\langle i,j \rangle \in\alpha\beta(\gamma)} & & K S_i^\gamma S_j^\gamma + \Gamma (S_i^\alpha S_j^\beta + S_i^\beta S_j^\alpha) \nonumber \\
& & + \Gamma' (S_i^\alpha S_j^\gamma + S_i^\gamma S_j^\alpha + S_i^\beta S_j^\gamma + S_i^\gamma S_j^\beta),
\end{eqnarray}
where $\langle i,j \rangle$ denotes nearest-neighbor sites, $\gamma$ labels the type of the bond $\langle i,j \rangle$ on the honeycomb lattice, and $\alpha,\beta,\gamma$ stand for the spin index. In most Kitaev materials the Kitaev terms have negative sign $K <0$ \cite{lukas,Jinsheng,rins,rwdyl,rcm}. In the present work, we adopt the parameters of the interactions such that $K<0$, $\Gamma>0$ and $\Gamma'$ is either positive or negative. Due to spin-orbit coupling, the symmetry of the model is described by the finite magnetic point group $D_{3d} \times Z_2^T$ besides lattice translation symmetries, where $Z_2^T = \{E,T\}$ is the time reversal group.

\begin{figure}[t]
\includegraphics[width=8.8cm]{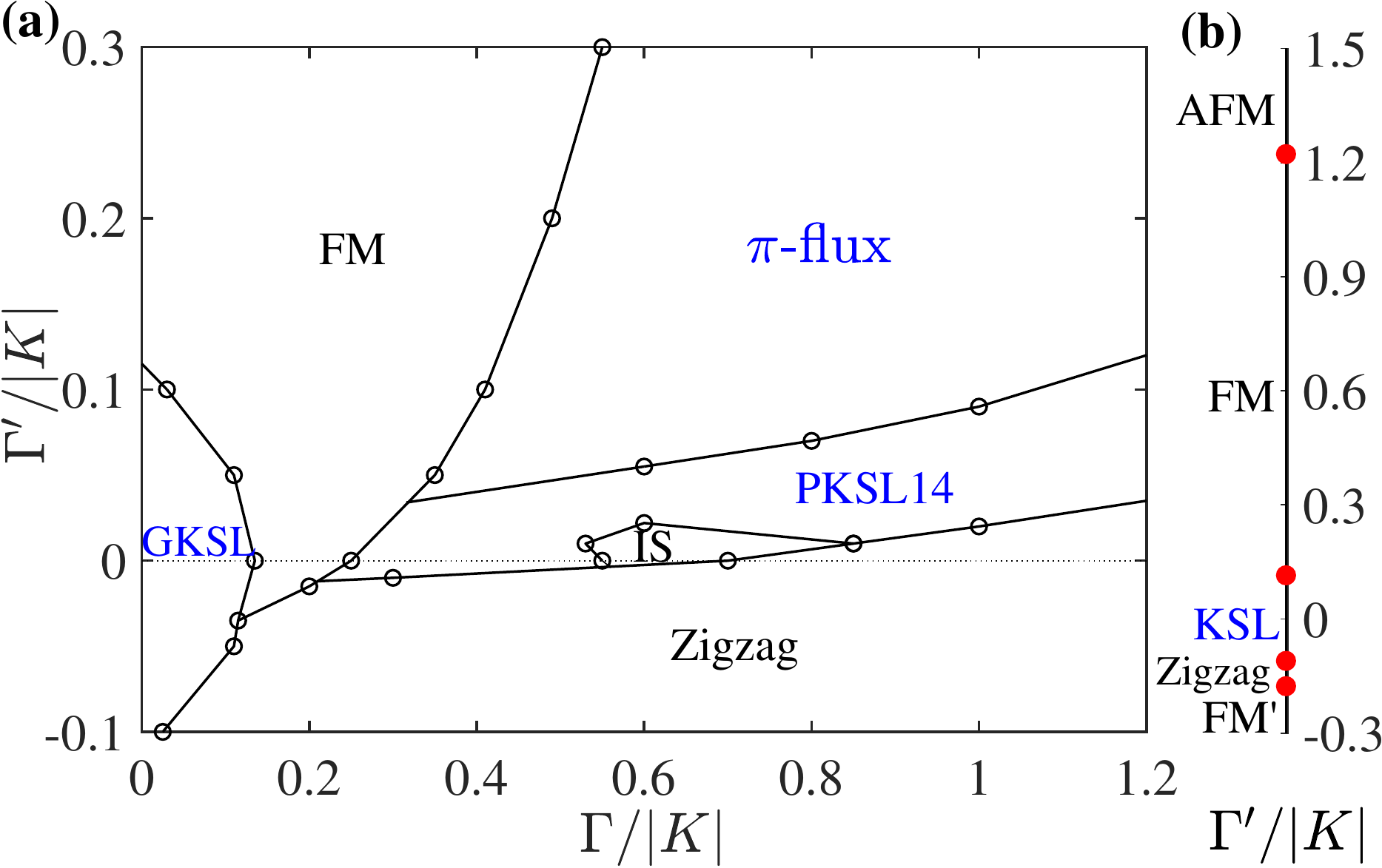}
\caption{ Phase diagrams of the quantum $K$-$\Gamma$-$\Gamma'$ model for $K < 0$, $\Gamma >0$ in the limit of large system size.
(a) Intermediate ${\Gamma/|K|}$ and ${\Gamma'/|K|}$, where three QSL phases (namely, the two-cone generic Kitaev spin liquid (GKSL) phase,  the 14-cone proximate Kitaev spin liquid phase (PKSL14), and the 32-cone $\pi$-flux state) and three magnetically ordered phases [including the ferromagnetic (FM) phase, the incommensurate spiral (IS) phase, and the zigzag phase] are found;
(b) The limit $\Gamma=0$, where the PKSL14 and $\pi$-flux phases vanish and two more ordered phases ({\it i.e.}, the AFM phase and the FM$'$ phase) appear. }
\label{KGammaGammaPrime}
\end{figure}


We study the model (\ref{KGG}) with the VMC method. Our calculations are performed on a  tori of up to 10$\times$10 unit cells, {\it i.e.}~of 200 lattice sites. We present the phase diagram here in Fig.~\ref{KGammaGammaPrime}, and leave the details of VMC calculations to Sec.~\ref{Gutzwiller}.

\begin{figure}
\includegraphics[width=4.2cm]{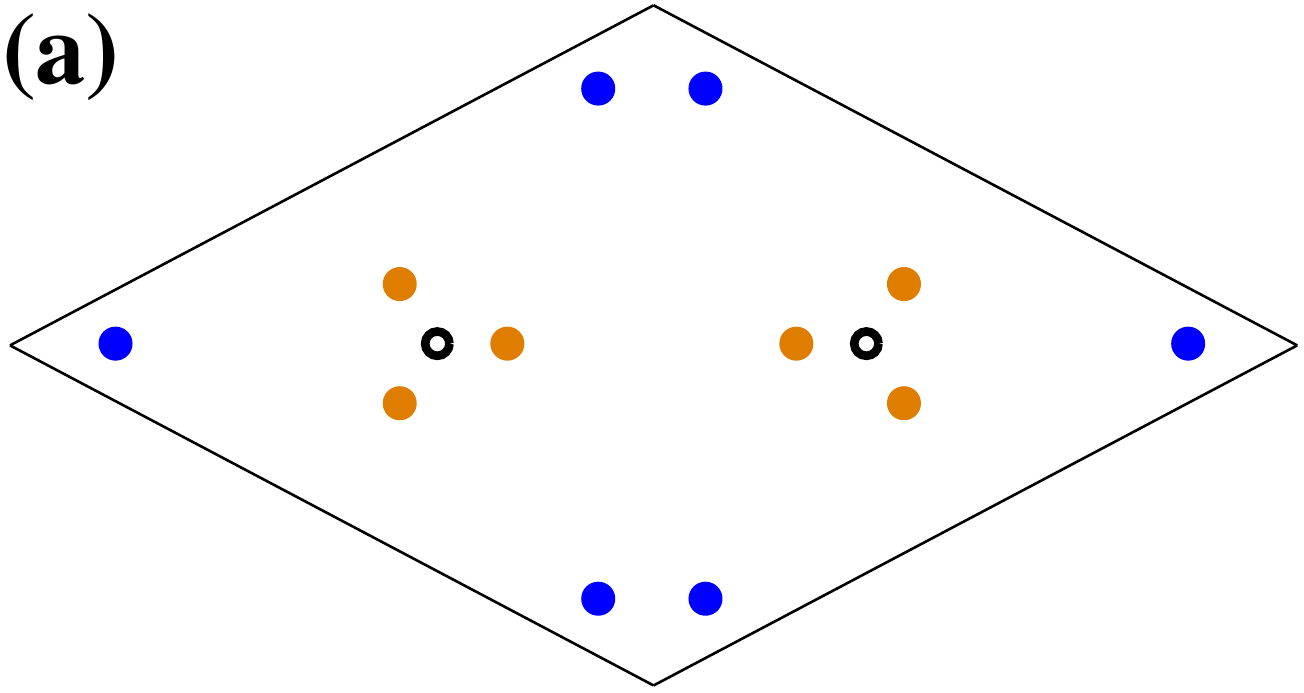}
\includegraphics[width=4.2cm]{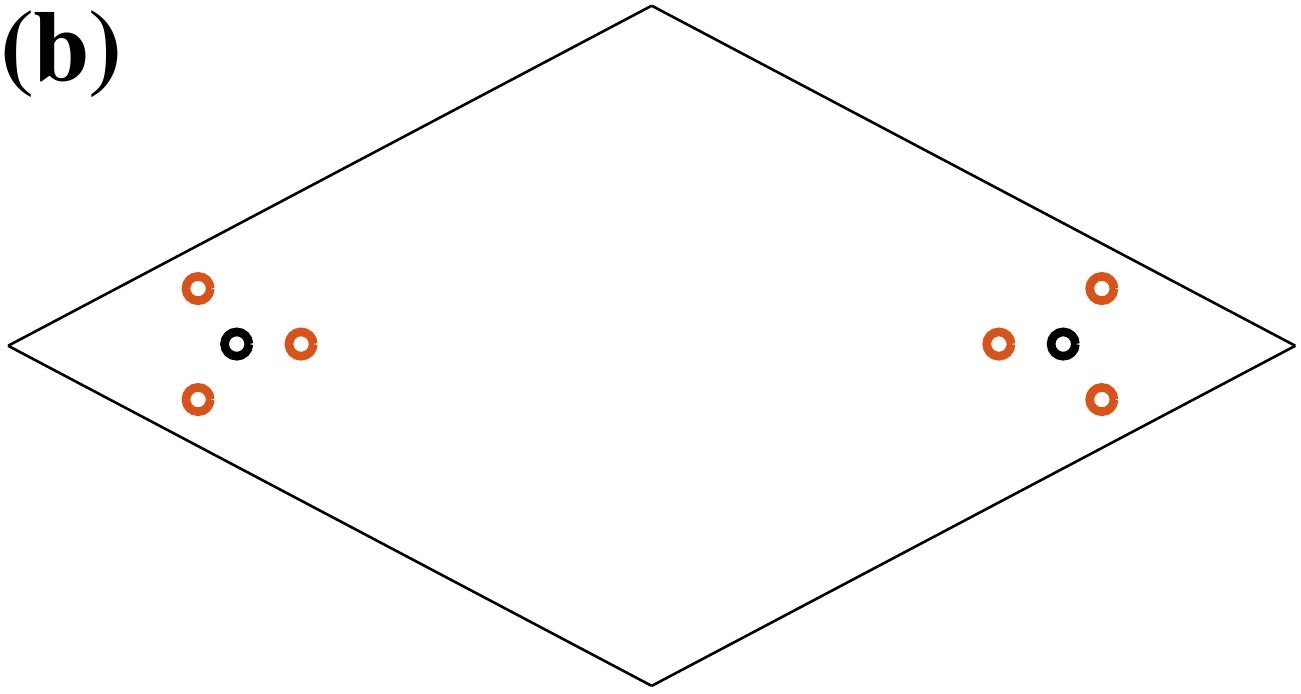}
\caption{Positions of the cones in the two gapless QSL phases. The dots in the same color are symmetry related. The solid dots stand for positive chirality while the hollow ones mean negative chirality. (a) The PKSL14 state with $\Gamma/|K|=1$ and $\Gamma'/|K|=0.05$. The rhombus is the conventional BZ, and the two dark hollow dots stand for $\pmb K={2\over3}\pmb G_1+{1\over3}\pmb G_2$ and $\pmb K'={1\over3}\pmb G_1+{2\over3}\pmb G_2$, respectively, which are invariant under $C_3$ rotation (up to a reciprocal lattice vector). (b) The $\pi$-flux state with $\Gamma/|K|=1$ and $\Gamma'/|K|=0.3$. The rhombus is the compact BZ (the magnetic BZ contains two copies of the compact BZ), and the two hollow dark dots stand for $\pmb K_1={5\over6} {\pmb G_1\over2}+{1\over6}{\pmb G_2\over2}$ and $\pmb K_1'={1\over6}{\pmb G_1\over2}+{5\over6} {\pmb G_2\over2}$, respectively. The $C_3$ and inversion symmetry of the excitation spectrum is discussed in Appendix \ref{app:cones}.
}
\label{Cones}
\end{figure}

Three different QSL phases are obtained. The phase containing the exactly solvable point is called the generic KSL (GKSL) phase whose spinon excitation spectrum contains two Majorana cones in the first BZ.  The GKSL is bounded approximately by $|\Gamma'|/|K|{} = 0.1$ at $\Gamma = 0$ and $\Gamma/|K|$ = 0.15 in Fig.~\ref{KGammaGammaPrime}(a). Another gapless QSL named PKSL14 locates at the regions $\Gamma/|K| > 0.2$ and $\Gamma'/|K|>-0.02$. This QSL phase shares the same PSG as the KSL and has 14 Majorana cones in its spinon excitation spectrum [see Fig.~\ref{Cones}(a) for the positions of the cones]. Finally, with the increase of $\Gamma'$, a $\pi$-flux state with a distinct PSG shows up which contains eight cones [see Fig.~\ref{Cones}(b) ] in the compact BZ. Later we will show that there are actually 32 cones in the original first BZ (see Sec.\ref{piflux}), which can be observed experimentally.


Besides the QSL phases, three magnetically ordered phases appear in Fig.~\ref{KGammaGammaPrime}(a), namely, the zigzag phase, the ferromagnetic (FM) phase, and the incommensurate spiral (IS) phase. The zigzag ordered phase is found to be robust when $\Gamma'/|K|<0$ and extends to large $\Gamma$ from $\Gamma/|K| \sim 0.15$, which is consistent with Refs.~\onlinecite{HYKee, tensor}.  The FM phase is bounded approximately by $\Gamma/|K| < 0.6$ and $\Gamma'/|K|>0$, while the small region sandwiched by the PKSL14 and the zigzag phase is the IS phase.

Figure \ref{KGammaGammaPrime}(b) shows the special case with $\Gamma=0$, where the system falls in the AFM phase ($\Gamma'\to+\infty$) or the FM$'$ phase ($\Gamma'\to-\infty$) in the large $|\Gamma'|$ limit. In this case, there is only one QSL phase (KSL). Therefore, the $\Gamma$ interaction is important to stabilize the non-Kitaev QSL phases.

All of the phase transitions between different phases in the phase diagram are of first order. The details of phase transitions are discussed in Appendix \ref{transitions}. The phase boundaries slightly depend on the size of the system. In Appendix \ref{finitesize}, we illustrate that in the large-size limit the phase boundaries are qualitatively the same as those in Fig.~\ref{KGammaGammaPrime}.


\section{Construction of Trial wave functions}\label{Gutzwiller}

Our VMC approach is based on spinon representation, where the spin operators are written in quadratic forms of fermionic spinons $S_i^m =\frac{1}{2} C_i^\dagger \sigma^m C_i$, where $C_i^\dagger = (c_{i\uparrow}^\dagger,c_{i\downarrow}^\dagger)$, $m \equiv x,y,z$, and $\sigma^m$ are Pauli matrices. The particle number constraint, $\hat{N_i} = c_{i\uparrow}^\dagger c_{i\uparrow} + c_{i\downarrow}^\dagger c_{i\downarrow} = 1$, should be imposed at every site such that the size of the Hilbert space of the fermions is the same as that of the original spin. It is convenient  to introduce the matrix operator $\psi_i=( C_i, \bar C_i)$ with $\bar C_i=(c_{i\dn}^\dag, -c_{i\up}^\dag)^T$ such that the spin operators can also be written as $S^m = {\rm Tr}(\psi_i ^\dag {\sigma^m \over4}\psi_i)$. In this form  it is easily  seen that there is a local $SU(2)$ gauge symmetry\cite{Anderson88} in the fermionic representation of spins.

The spin interactions in Eq.~(\ref{KGG}) are rewritten in terms of interacting fermionic operators and are further decoupled into a non-interacting mean-field Hamiltonian $H_{\rm mf}(\pmb R)$, where $\pmb R$ denotes a set of parameters and will be specified in later discussion.

We perform Gutzwiller projection to the mean-field ground state $|\Psi_{\rm mf} (\pmb R)\rangle$ to ensure the particle  number constraint. The projected states $|\Psi (\pmb R)\rangle = P_G |\Psi_{\rm mf}(\pmb R) \rangle$ provide a series of trial wave functions depending on the choice of the mean-field Hamiltonian $H_{\rm mf}(\pmb R)$, where $P_G$ denotes a Gutzwiller projection and $\pmb R$ are treated as variational parameters.  The energy of the trial state $E (\pmb R) = \langle \Psi(\pmb R) |H| \Psi(\pmb R) \rangle / \langle \Psi(\pmb R)| \Psi(\pmb R) \rangle$ is computed using Monte Carlo sampling, and the optimal parameters $\pmb R$ are determined by minimizing the energy $E(\pmb R)$.

While the VMC calculations are performed on a relatively small size (up to $200$ sites), once the optimal parameters are determined we can plot the spinon dispersion of a QSL state by diagonalizing the mean-field Hamiltonian on a larger lattice size (we adopted $120\times120$ unit cells).

\subsection{Spin-liquid ansatzes based on PSG}\label{SM:MFs}

A QSL ground state preserves the whole space group symmetry whose point group is $D_{3d}\times Z_2^T$. However, the symmetry group of a spin liquid mean-field Hamiltonian is the projective symmetry group (PSG) \cite{igg,You_PSG} whose group elements are space group operations followed by $SU(2)$ gauge transformations.

It  turns out that there are more than 100 classes of PSGs for Z$_2$ QSLs [in a $Z_2$ QSL the $SU(2)$ "gauge symmetry" breaks down to the $Z_2$ subgroup in the mean-field Hamiltonian] respecting the $D_{3d}\times Z_2^T$ symmetry\cite{You_PSG}. It is not practical to investigate all of these PSGs. In our VMC calculations, we consider a few of them (all in class I-B or class I-A) which are close to the one which describes the symmetry of the exact ground state of the pure Kitaev model.
Here "close" means that the new PSGs and the Kitaev PSG have similar patterns of symmetry fractionalization, namely, they differ by only one or two invariants. The reason for choosing these PSGs is based on the fact that the model (\ref{KGG}) contains Kitaev interactions. The ground state of the pure Kitaev model belongs to the Kitaev PSG class. According to our calculation, the existence of Kitaev interactions in our model is essential to stabilize the QSL phases (similar conclusions also appeared in the literature). Therefore, it is reasonable to adopt the PSGs that are "close to the Kitaev PSG" given that the non-Kitaev interactions are not extremely large.

Now we provide details of constructing $H_{\rm mf}$ for a given PSG. The most general expression of the mean-field Hamiltonian ansatz\cite{PKSL,Liu_KG,Aniso}  with nearest neighbor couplings reads (as shown in appendix \ref{NNNC}, the next-nearest-neighbor couplings and further long-range coupling terms are not important),
\beq\label{MFPSG}
H_{\rm mf}^{\rm SL} & = & \!\!\!\! \sum_{\langle i,j \rangle \in\alpha\beta(\gamma)} \!\!\! {\rm Tr}
\, [U_{ji}^{(0)} \psi_i^\dag \psi_j] \! + \! {\rm Tr} \, [U_{ji}^{(1)} \psi_i^\dag
(i R_{\alpha\beta}^\gamma) \psi_j] \nonumber\\ & & \;\;\;\; + {\rm Tr} \, [U_{ji}^{(2)}
\psi_i^\dag \sigma^\gamma \psi_j] \! + \! {\rm Tr} \, [U_{ji}^{(3)} \psi_i^\dag
\sigma^\gamma R_{\alpha\beta}^\gamma \psi_j] \! + \! {\rm H.c.} \nonumber\\
& & \;\;\;\; +  \sum_i {\rm Tr}  (\pmb \lambda_i\cdot  \psi_i \pmb \tau\psi_i^\dag ),
\eeq
where $R_{\alpha\beta}^\gamma = - \frac{i}{\sqrt{2}} (\sigma^\alpha + \sigma^\beta)$,  $\lambda^{x,y,z}$ are three Lagrangian multipliers to ensure SU(2) gauge invariance (where $\lambda^z$ is the one for the particle number constraint), $\tau^{x,y,z}$ are generators of the SU(2) gauge group, and the matrices $U_{ji}^{(0,1,2,3)}$ can be expanded with the identity matrix and $\tau^{1,2,3}$ where the expanding coefficients form a subset of $\pmb R$.  Generally, the values of $\lambda^{x,y,z}$ are zero if there are no external magnetic fields. Therefore, we only need to consider $\lambda^{x,y,z}$ when $\pmb B\neq0$ [see Eq.~(\ref{Bfield})].

As shown in the following, the PSG constrains the values of the matrices $U_{ji}^{(0,1,2,3)}$.



\subsubsection{ The Kitaev PSG [class (I-B) $0$-flux]} 
The gapless KSL is believed to be a finite, stable phase in the presence of non-Kitaev interactions, including the $\Gamma$ and $\Gamma'$ terms.  The mean-field Hamiltonian describing the generic states around the KSL, which we denote the GKSL, will then respect the same PSG as the KSL itself. Besides translation symmetry, the symmetry group of the pure KSL, $G = D_{3d}$$\times$$Z_2^T$, has the three generators
\[
S_6 = (C_3)^2 P,\ \ \  M = C_2^{x-y} P,\ \ \   T = i\sigma^y K,
\]
where $C_3$ is a threefold rotation around the direction $\hat c \equiv {\textstyle
\frac{1}{\sqrt3}} (\hat x + \hat y + \hat z)$, $C_2^{x-y}$ is a twofold rotation
around ${\textstyle \frac{1}{\sqrt2}} (\hat x-\hat y)$, and $P$ is spatial
inversion. The PSG of the KSL (called Kitaev PSG) is read most simply from the Majorana
representation, in which the mean-field Hamiltonian is
\begin{eqnarray}\label{Kitaevmf}
H_{\rm mf}^{K} & = & \sum_{\langle i,j \rangle \in\alpha\beta(\gamma)} \rho_a (ic_ic_j) +
\rho_c (i b_i^\gamma b_j^\gamma)  \\ & = & \sum_{\langle i,j \rangle \in\alpha\beta(\gamma)}
i \rho_a {\rm Tr} \left(\psi_i^\dagger \psi_j + \tau^x \psi_i^\dagger
\sigma^x \psi_j + \tau^y \psi_i^\dagger \sigma^y \psi_j \right. \nonumber \\ & &
\;\;\;\;\;\;\; \left. + \tau^z \psi_i^\dagger \sigma^z \psi_j \right) + i \rho_c {\rm Tr}
\left(\psi_i^\dagger \psi_j + \tau^\gamma \psi_i^\dagger \sigma^\gamma \psi_j \right. \nonumber
\\ & & \;\;\;\;\;\; \left. - \tau^\alpha \psi_i^\dagger \sigma^\alpha \psi_j - \tau^\beta
\psi_i^\dagger \sigma^\beta \psi_j\right) + {\rm H.c.} \nonumber
\end{eqnarray}
Because the $c$ fermion never mixes with any of the $b^m$ fermions, any PSG
operation leaves the $c$ fermions invariant. The gauge operation, $W_i(g)$,
following the symmetry operation $g$ should then be $W_i(g) = \pm g$.
A detailed analysis \cite{You_PSG,PKSL} shows that the gauge transformations of the generators $S_6$, $M$, and $T$ are
\beq\label{PSGKSL}
W_A(S_6) & =& - W_B(S_6) = \exp \! \big[-i {\textstyle \frac{4}{3}} \pi
{\textstyle \frac{1}{2\sqrt{3}}} (\tau^x + \tau^y + \tau^z) \big],
\nonumber \\
W_A(M) & =& - W_B(M) = \exp \! \big[-i \pi {\textstyle \frac{1}{2\sqrt{2}}}
(\tau^x - \tau^y) \big], \nonumber\\
W_A(T) & = &- W_B(T) = i \tau^y,
\eeq
where A and B denote the two sublattices of the honeycomb lattice.

When the Kitaev model is extended to the $K$-$\Gamma$-$\Gamma'$ model,
there are several different ansatzes for states beyond the Kitaev
mean-field Hamiltonian [Eq.~(\ref{Kitaevmf})] that are invariant under the
same PSG. The $\Gamma$ interaction gives rise to the mean-field terms
\beq\label{Gm}
H_{\rm mf}^{\Gamma} & = & \!\! \sum_{\langle i,j \rangle \in\alpha\beta(\gamma)} \!\! i\rho_d
(b_i^\alpha b_j^\beta + b_i^\beta b_j^\alpha) \\ & = & \!\! \sum_{\langle i,j \rangle \in\alpha\beta(\gamma)} \!\! i\rho_d {\rm Tr} \left( \tau^\alpha \psi_i^\dag
\sigma^\beta \psi_j + \tau^\beta \psi_i^\dag \sigma^\alpha \psi_j\right)
 + {\rm H.c.} \nonumber
\eeq
and similarly for the $\Gamma'$ interaction
\beq\label{Gmp}
H_{\rm mf}^{\Gamma'} & = & \!\! \sum_{\langle i,j \rangle \in\alpha\beta(\gamma)} \!\! i\rho_f
(b_i^\alpha b_j^\gamma + b_i^\gamma b_j^\alpha + b_i^\beta b_j^\gamma + b_i^\gamma b_j^\beta) \\
& = & \!\! \sum_{\langle i,j \rangle \in\alpha\beta(\gamma)} \!\! i\rho_f {\rm Tr} \left( \tau^\alpha \psi_i^\dag
\sigma^\gamma \psi_j + \tau^\gamma \psi_i^\dag
\sigma^\alpha \psi_j \right. \nonumber \\
&  &\phantom{=\;\;} \left. + \tau^\beta \psi_i^\dag \sigma^\gamma \psi_j + \tau^\gamma \psi_i^\dag \sigma^\beta \psi_j \right) +  {\rm H.c.}  \nonumber
\eeq

Comparing with  the general form Eq.~(\ref{MFPSG}),  the decouplings expressed in
Eqs.~(\ref{Kitaevmf}), (\ref{Gm}) and (\ref{Gmp}) contribute the terms
\begin{equation}
\begin{aligned}
& {\tilde U}_{ji}^{(0)} = i (\rho_a + \rho_c), \\
& {\tilde U}_{ji}^{(1)} = i (\rho_a - \rho_c  + \rho_d + 2\rho_f) (\tau^\alpha + \tau^\beta),
\\ & {\tilde U}_{ji}^{(2)} = i (\rho_a + \rho_c) \tau^\gamma + i \rho_f(\tau^\alpha +\tau^\beta ), \\
& {\tilde U}_{ji}^{(3)} = i (\rho_c - \rho_a - \rho_d) (\tau^\alpha - \tau^\beta ),
\end{aligned}
\end{equation}
to the coefficients $U_{ji}^{(m)}$, in which $j$ and $i$ specify $\gamma$.
However, the most general coefficients preserving the $C_3$ rotation symmetry
(in the PSG sense) also contain multiples of the uniform ($I$) and $\tau^x +
\tau^y + \tau^z$ gauge components,
\begin{equation}
\begin{aligned}
& {\tilde {\tilde U}_{ji}^{(0)}} = i \eta_0 + \eta_1 (\tau^x + \tau^y + \tau^z), \\
& {\tilde {\tilde U}_{ji}^{(1)}} = \eta_2 + i \eta_3 (\tau^x + \tau^y + \tau^z), \\
& {\tilde {\tilde U}_{ji}^{(2)}} = \eta_4 + i \eta_5 (\tau^x + \tau^y + \tau^z), \\
& {\tilde {\tilde U}_{ji}^{(3)}} = \eta_6 + i \eta_7 (\tau^x + \tau^y + \tau^z).
\end{aligned}
\end{equation}
If the full symmetry group, $G = D_{3d} \times Z_2^T$, is preserved, then only three parameters $\eta_0$, $\eta_3$, and $\eta_5$ are allowed; by contrast, if one allows the breaking of spatial inversion symmetry, while still preserving mirror reflection symmetry, then $\eta_1$, $\eta_2$, and $\eta_4$ are also allowed. Thus a spin-liquid ansatz that preserves the full PSG symmetry generated by Eq.~(\ref{PSGKSL}) contains the variables $U_{ji}^{(m)} = {\tilde U}_{ji}^{(m)} + {\tilde {\tilde U}_{ji}^{(m)}}$ with seven real parameters, $\rho_a$, $\rho_c$, $\rho_d$, $\rho_f$, $\eta_0$, $\eta_3$ and $\eta_5$. Of these, only $\eta_3$ and $\eta_5$ lead to a hybridization of the $c$ with the $b^m$ fermions, which has important consequences for the spin response of the ground state.

The GKSL, PKSL14, PKSL8$_2$, and PKSL8$_4$ states belong to the Kitaev PSG class.

\subsubsection{ Class (I-B) $\pi$-flux PSG}\label{sec:piflux}
As a special example in class (I-B), we give the ansatz in which the spinons feel a uniform $\pi$-flux on each hexagon called the $\pi$-flux state (shown in Fig.\ref{PSGs}). The general form preserving the full symmetry group reads $U_{ji}^{(m)}  = (-\tau^0)^{ji} ({\tilde U}_{ji}^{(m)}  + {\tilde {\tilde U}_{ji}^{(m)}} )$.
We use $(-\tau^0)^{ji}$ to note the sign pattern of the uniform $\pi$-flux in each hexagon with double unit cell (for details see Appendix \ref{BZ}).
Therefore, the $\pi$-flux state also contains seven variational parameters which are similar to the above ansatz.


\subsubsection{ Another class (I-B) $0$-flux PSG}

As another example in class (I-B), we give the ansatz with $\eta_1=\eta_9=-\eta_{10}=-\eta_{14}=\tau^0$ and $\theta_1=2\pi/3$ (see Ref.~\onlinecite{You_PSG} for details), namely the SL-B state (shown in Fig.\ref{PSGs}).
The general form preserving the full symmetry group reads $U_{ji}^{(m)}  = {\tilde {\tilde U}_{ji}^{(m)}} $.
If the full symmetry group is preserved, then
only four parameters $\eta_1$, $\eta_2$, $\eta_4$, and $\eta_6$ are allowed.


\subsubsection{ Class (I-A) PSG}
As one example in class (I-A), we give the ansatz with $\eta_1=\eta_9=-\eta_{10}=-\eta_{14}=\tau^0$ (see Ref.~\onlinecite{You_PSG} for details), namely the SL-A state (shown in Fig.\ref{PSGs}).
If the full symmetry group is preserved, then coefficients $U_{ji}^{(m)}$ read
\begin{equation}
\begin{aligned}
& U_{ji}^{(0)} = \phi_1(\tau^x-\tau^y), \\
& U_{ji}^{(1)} = i\phi_2 \tau^z, \\
& U_{ji}^{(2)} = i\phi_3\tau^z, \\
& U_{ji}^{(3)} = i\phi_4 (\tau^x - \tau^y).
\end{aligned}
\end{equation}
Therefore, only four parameters $\phi_1$, $\phi_2$, $\phi_3$, and $\phi_4$ are allowed.


\subsection{Magnetically ordered states}
To describe the magnetic order of the spin-symmetry-breaking phases of the $K$-$\Gamma$-$\Gamma'$ model, we introduce the classical order under single-${\pmb Q}$ approximation\cite{singleQ}
\[\pmb{M}_i = M \{\sin \phi [\hat {\pmb e}_x \cos (\pmb{Q} \cdot
\pmb{r}_i) + \hat{\pmb e}_y \sin(\pmb{Q} \cdot \pmb{r}_i)] + \cos \phi
\, \hat{\pmb e}_z \},\]
where  $\pmb Q$ is the ordering momentum, $\hat {\pmb e}_{x,y,z}$ are the local spin axes (not to be confused with the global spin axes),
and $\phi$ is the canting angle. $\pi/2-\phi$ describes the angle by which the spins deviate from the plane spanned by $\hat {\pmb e}_x$ and $\hat {\pmb e}_y$. The classical ground state is obtained by minimizing the energy of the trial states.

In our VMC calculations, the static order is treated as a background field coupling to the spins as site-dependent Zeeman field; 
hence the complete mean-field Hamiltonian for the $K$-$\Gamma$-$\Gamma'$ model reads
\begin{equation}\label{Order}
H_{\rm mf}^{\rm total} = H_{\rm mf}^{\rm SL} - {\textstyle \frac{1}{2}} \sum_i
(\pmb {M}_i \cdot C_i^\dagger \pmb \sigma C_i + {\rm H.c.})
\end{equation}

The ordering momentum $\pmb Q$ of $\pmb M_i$ in VMC is adopted from the classical ground state or the classical metal stable states (depending on the energy of the projected state). For a given $\pmb Q$, the local axes $\hat {\pmb e}_{x,y,z}$ are fixed as they are in the classical state,  $M$ and $\phi$ are treated as variational parameters. 
In the ordered ground states except for those in the IS phase, the canting angle $\phi$ is very close to $\pi/2$.
Thus the quantum corrections we compute are in essence contained in the amplitude of the static magnetic order $M$.

In our VMC calculation, we have also considered the six-site magnetic order proposed in Refs.~\onlinecite{tensor} and \onlinecite{class}, which is beyond the single-$\pmb Q$ approximation. To include this order, we consider a system with $6\times4$ unit cells where each unit cell contains six sites. However, the ordering magnitude of our VMC outputs is vanishingly small ($M\approx 0.01$). This indicates that the six-site order is not favored by the $K$-$\Gamma$-$\Gamma'$ model in the VMC approach.

\subsection{VMC-selected ground states}\label{VMCgs}


In order to provide a complete and systematic study of the $K$-$\Gamma$-$\Gamma'$
phase diagram, we have investigated many different ansatzes. 
Here we only consider the parameter interval where magnetically ordered states are not favored in energy. We calculate the energy of the projected 
ansatz given in Sec.\ref{SM:MFs}; 
the one with the lowest energy is treated as the ground state of the system.

\begin{figure}[t]
\includegraphics[width=8.5cm]{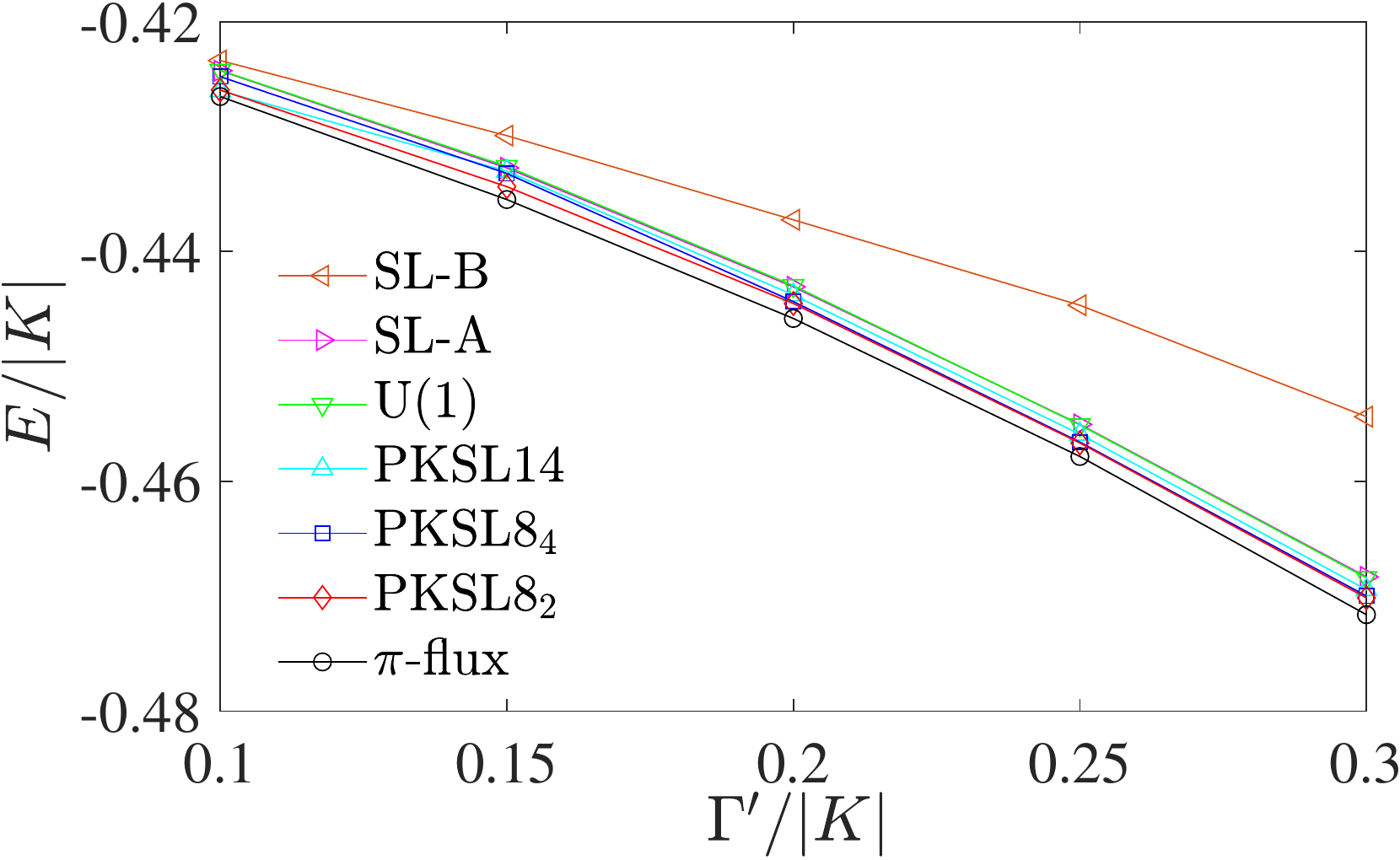} \
\caption{Ground-state energy per site, $E$, of the $K$-$\Gamma$-$\Gamma'$ model ($\Gamma/|K|
 = 0.8$), comparing the lowest-lying trial wave functions for U(1) and Z$_2$ QSLs.}
\label{PSGs}
\end{figure}

To illustrate the competition between different ansatz, in Fig.~\ref{PSGs} we show the energy curves of various trial states for fixed  $\Gamma/|K| =0.8 $. It is clearly seen that  the $\pi$-flux state is lowest in energy in the parameter region of $0.1<\Gamma'/|K|<0.3$. However, two kinds of PKSL8 states, namely PKSL8$_2$ and PKSL8$_4$, are competing in energy with the $\pi$-flux state and will be discussed in more detail in Sec.\ref{family}.

%
%

\begin{figure}[t]
\includegraphics[width=8.3cm]{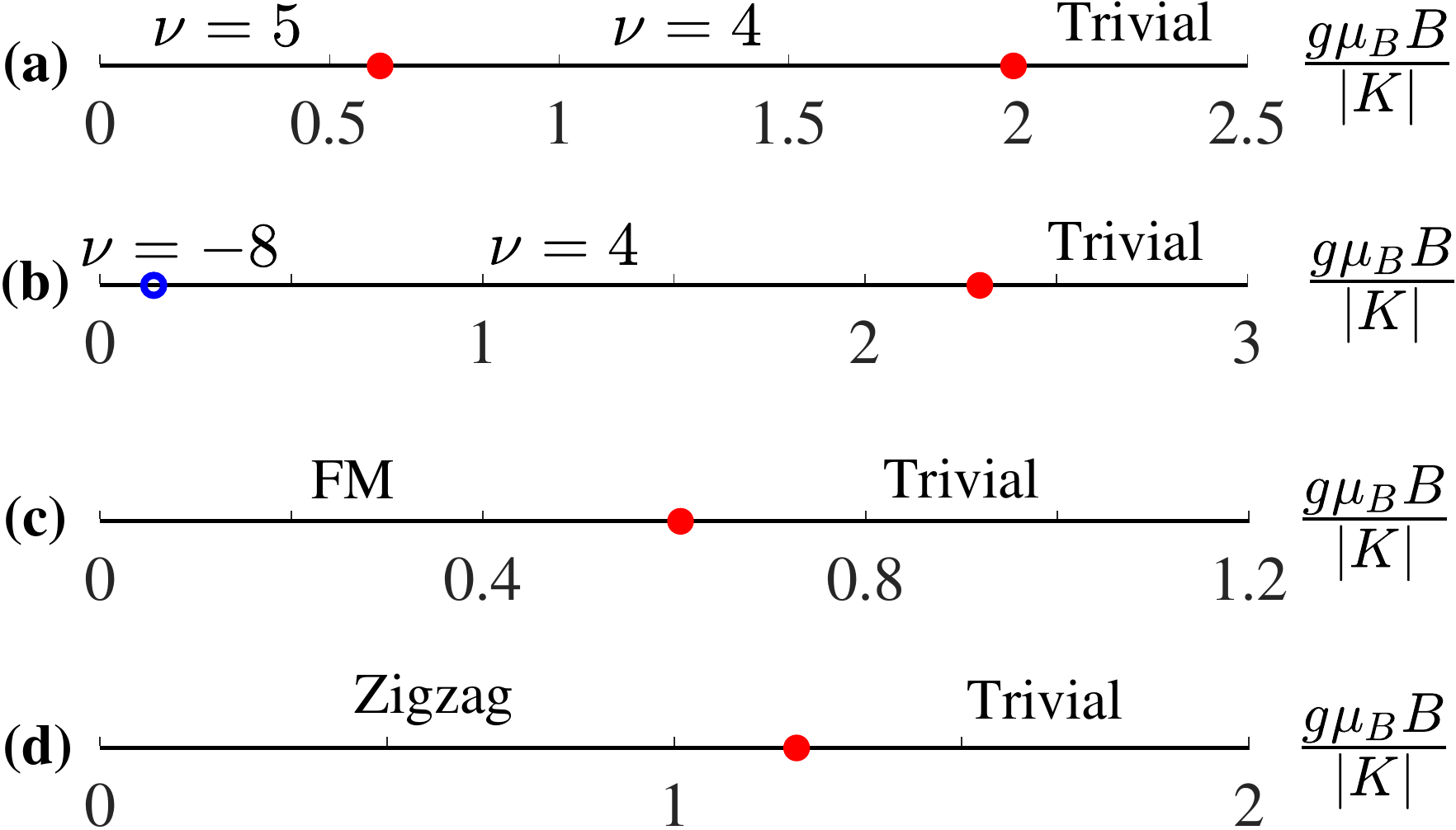} \
\caption{Phase diagrams when a magnetic field is applied with $\pmb B\parallel(\hat x+ \hat y+ \hat z)$.
(a) At $\Gamma/|K|=1$, $\Gamma'/|K|=0.05$, the PKSL14 evolves into one non-Abelian CSL ($\nu=5$) and one Abelian CSL ($\nu=4$).
(b) At $\Gamma/|K|=1$, $\Gamma'/|K|=0.3$, the $\pi$-flux state is turned into two gapped Abelian CSLs ($\nu=-8, 4$).
(c) At $\Gamma/|K|=0.2$, $\Gamma'/|K|=0.1$, a direct first-order phase transition occurs from the FM phase to the trivial phase.
(d) At $\Gamma/|K|=1.4$, $\Gamma'/|K|=-0.05$, a direct first-order phase transition occurs from the zigzag phase to the trivial phase.
The red solid (blue hollow) points represent first-order (continuous) phase transitions. }
\label{Bc}
\end{figure}

\subsection{More about the $\pi$ flux phase}\label{piflux}
Now we come to a previously mentioned subtle issue, namely the number of cones in the $\pi$-flux phase.
Although the mean-field Hamiltonian of the $\pi$-flux QSL has a doubled unit cell and halved magnetic BZ, the dispersion of the spinon bands has a twofold ``translation degeneracy'' \cite{fluxcrystal} in the original BZ along both $\pmb G_{1}$  and $\pmb G_2$ directions (here  $\pmb G_{1},\pmb G_2$ are the reciprocal lattice vectors for the original lattice, see Appendix \ref{app:cBZ} for details). Thus the minimal unit of the dispersion is the compact BZ [see Figs.~\ref{Cones}(b) and \ref{lattice}(b)] spanned by $\frac{1}{2}\pmb G_{1},\frac{1}{2}\pmb G_{2}$, which contains eight cones. However, after Gutzwiller projection, the translation symmetry in the $\pi$-flux state is restored. Therefore, we should go back to the original BZ to observe the physical quantities. As mentioned above, the spinon dispersion in the original BZ has a twofold periodic structure in both $\pmb G_1$ and $\pmb G_2$ directions, so there are totally 32 cones. The periodic structure is an observable property of the $\pi$-flux state \cite{igg}.

In the present work, when we count the number of cones that are detectable in experiments, we adopt the original BZ; when we calculate the Chern number, we use the magnetic BZ; but when we illustrate the dispersion [see Fig.~\ref{Cones}(b)], we adopt the compact BZ.

\section{The effect of magnetic fields}\label{fields}

In this part, we consider the consequences of adding an external magnetic field to the gapless QSLs. To this end, we add a Zeeman term $H_{B}$ to $H_{\rm mf}$,
\beq\label{Bfield}
H_{B} = g\mu_B \sum_i {\rm Tr}  (\pmb B\cdot  \psi_i^\dag {\pmb \sigma\over2}\psi_i),
\eeq
Here we have ignored the anisotropy in the $g$ factor. We first consider the case $\pmb B\parallel (\hat x+ \hat y+ \hat z)$. It is known that in such a field the KSL opens a gap and becomes a non-Abelian CSL with Chern number $\nu=1$, where the non-Abelian statistics arise due to unpaired Majorana zero modes associated with the vortices \cite{Kitaev}. With the increasing strength of the field, the system undergoes a continuous phase transition (with a gap closing at $\pmb k=0$ point) to a trivial polarized phase.

\begin{figure}[t]
\includegraphics[width=8.3cm]{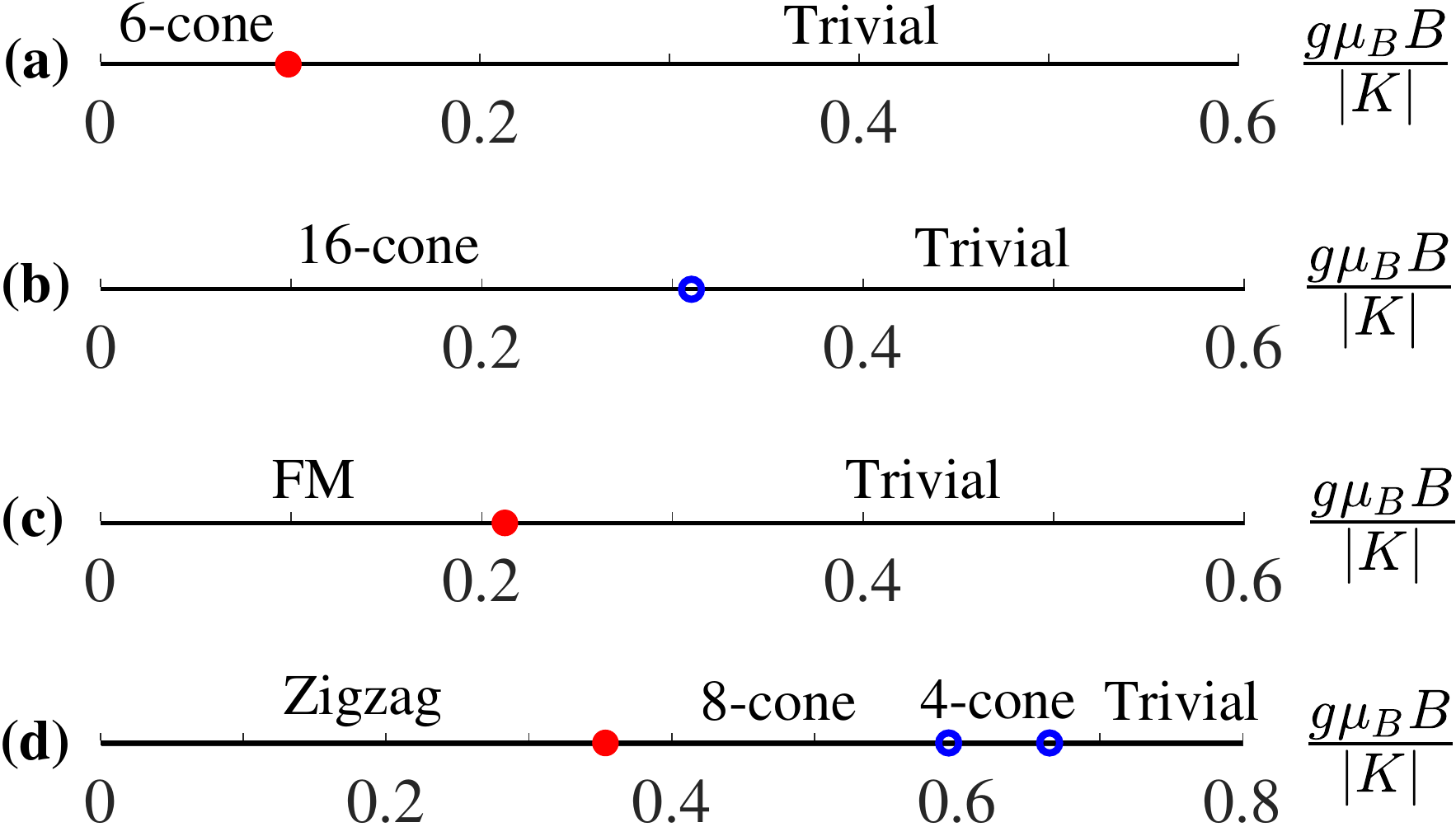} \
\caption{Phase diagrams when a magnetic field is applied with $\pmb B\parallel(\hat x- \hat y)$.
(a) At $\Gamma/|K|=1$, $\Gamma'/|K|=0.05$, the PKSL14 is turned into a 6-cone phase before entering the trivial phase.
(b) At $\Gamma/|K|=1$, $\Gamma'/|K|=0.3$, the $\pi$-flux state is turned into a 16-cone phase before being polarized by the field.
(c) At $\Gamma/|K|=0.2$, $\Gamma'/|K|=0.1$, a direct first-order phase transition occurs from the FM phase to the trivial polarized phase.
(d) At $\Gamma/|K|=1.4$, $\Gamma'/|K|=-0.05$), as the zigzag order is suppressed by the field, two gapless states (8 cones and 4 cones) are induced. The red solid (blue hollow) points represent first-order (continuous) phase transitions.
}
\label{Bxy}
\end{figure}

The 14 cones in the PKSL14 state can be divided into three groups which are marked by different colors in Fig.~\ref{Cones}(a). The cones within each group are symmetry-related (namely, they can be transformed into each other via symmetry operations), while the ones in different groups are independent. When magnetic field $\pmb B\parallel (\hat x+ \hat y+ \hat z)$ is applied, all of the cones are gapped out and the ones in each group contribute the same amount (namely, either $1\over2$ or $-{1\over2}$) to the total Chern number when the magnetic field is weak enough to be treated as a perturbation. Therefore, the total Chern number should be $\nu=3\chi_1+3\chi_2+\chi_3$, where $\chi_{1,2,3}=\pm 1$ denote the {\it chiralities} of the three groups of cones. Our numerical calculation indicates that $\chi_1=\chi_2=1,\chi_3=-1$ [see Fig.~\ref{Cones}(a), where the hollow (solid) dots stand for the cones with negative (positive) chirality] and thus the total Chern number is $\nu=5$. With increasing $|\pmb B|$, the system undergoes first-order phase transitions from the $\nu=5$ CSL to a $\nu=4$ CSL phase and then to the trivial polarized phase \cite{PKSL}.

Similarly,  a small $|\pmb B|$ turns the $\pi$-flux state into an Abelian CSL phase with $\nu = -8$, as shown in Fig.~\ref{Bc}(b) (noticing that the magnetic BZ contains two copies of compact BZs). Interestingly, a continuous phase transition is observed from the $\nu=-8$ CSL to the $\nu=4$ CSL at the critical field ${g\mu_B B/|K|}=0.14$. This continuous transition with Chern number changing by 12 is protected by the remaining $C_3$ and inversion symmetry in the presence of the field. Finally, the system enters the trivial phase at ${g\mu_B B/|K|}=2.3$ with a first-order transition.

The Chern numbers of the CSLs can be measured by their quantized thermal Hall conductance $\kappa_{xy} = {\nu\over2}(\pi k_B^2 T/ 6h)$ at low temperatures. All of the obtained CSLs belong to the Kitaev's 16-fold classification, where the ones with odd $\nu$ are non-Abelian and the ones with even $\nu$ are Abelian.

Then we consider the effects of in-plane magnetic fields, especially the case $\pmb B\parallel(\hat x- \hat y)$.  In the two gapless QSLs, the cones on the high symmetry line ({\it i.e.}, the horizontal line in the original BZ or the compact BZ) remain gapless while other cones are gapped out. For the PKSL14 state, before entering the polarized phase with a first-order transition, a  6-cone gapless phase is obtained [see Fig.~\ref{Bxy}(a)] which is much more robust compared with the PKSL phase at $\Gamma'=0$ \cite{PKSL}.  Similarly, a 16-cone gapless QSL is induced from the $\pi$-flux state by a weak field [see Fig.~\ref{Bxy}(b)]. With the increasing of field strength, the system undergoes a continuous transition from the 16-cone phase to the trivial gapped phase, where the cones merge in pairs and disappear simultaneously.

\begin{figure}[t]
\includegraphics[width=4.2cm]{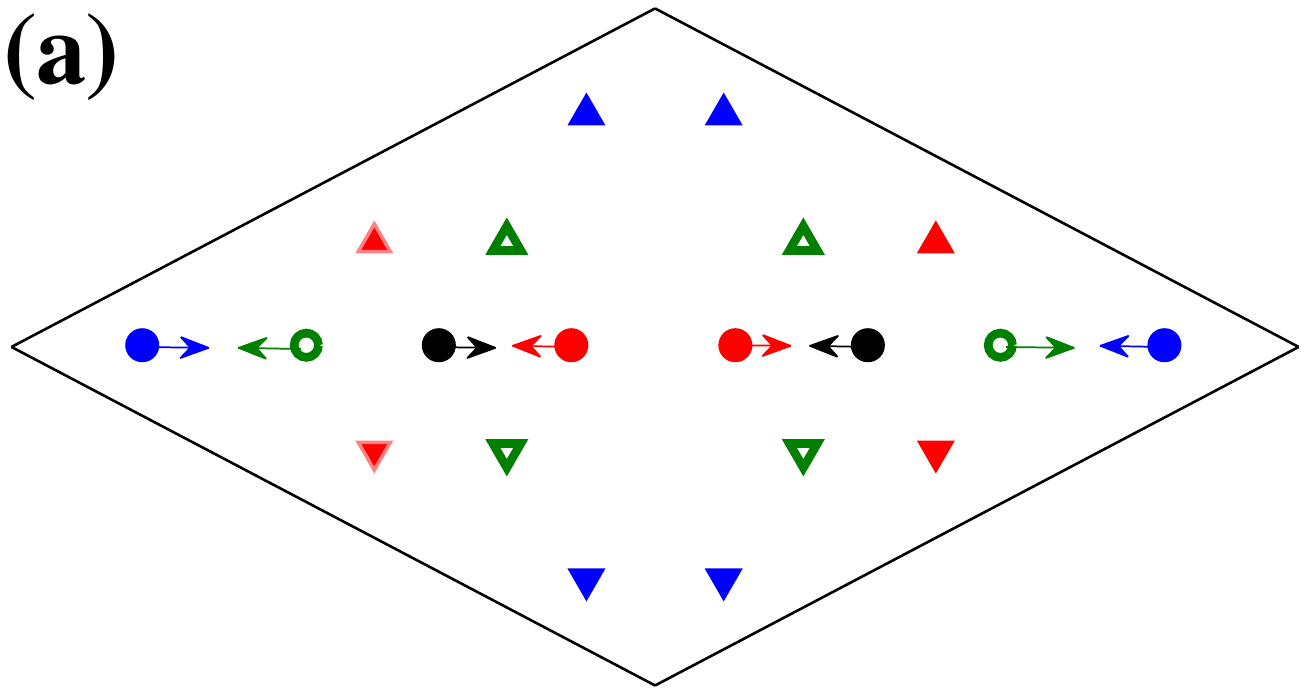}
\includegraphics[width=4.2cm]{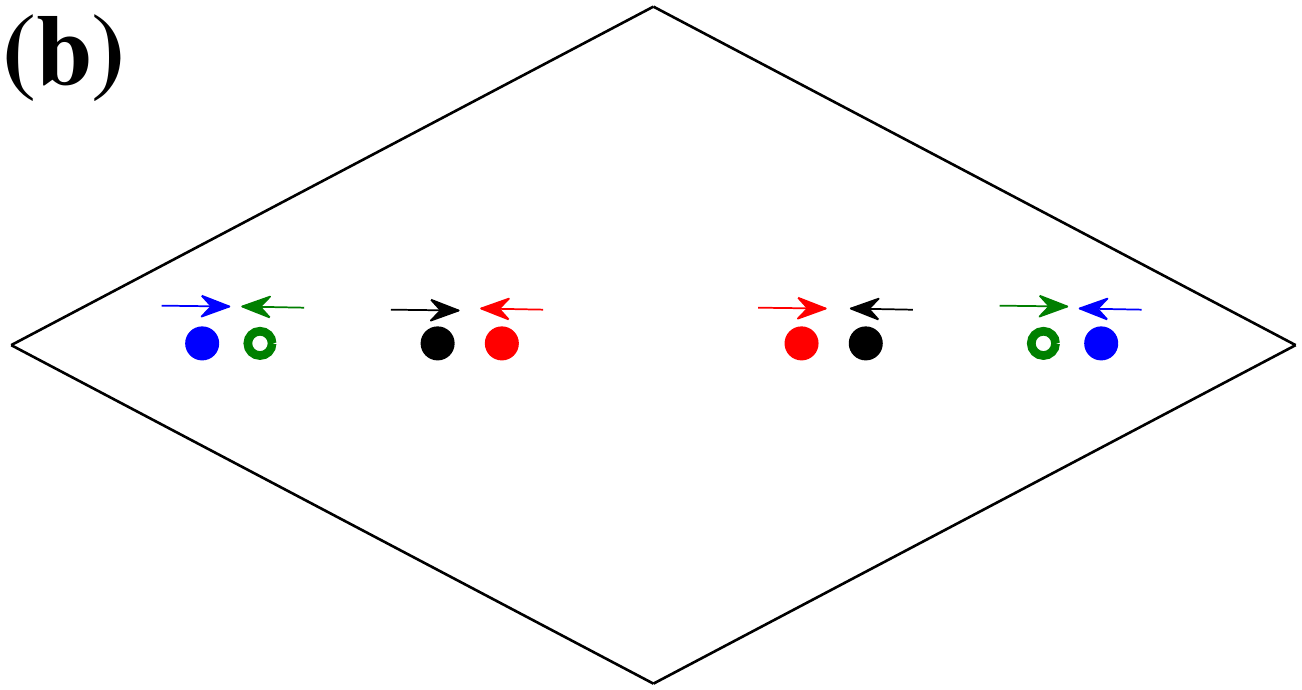}
\includegraphics[width=4.2cm]{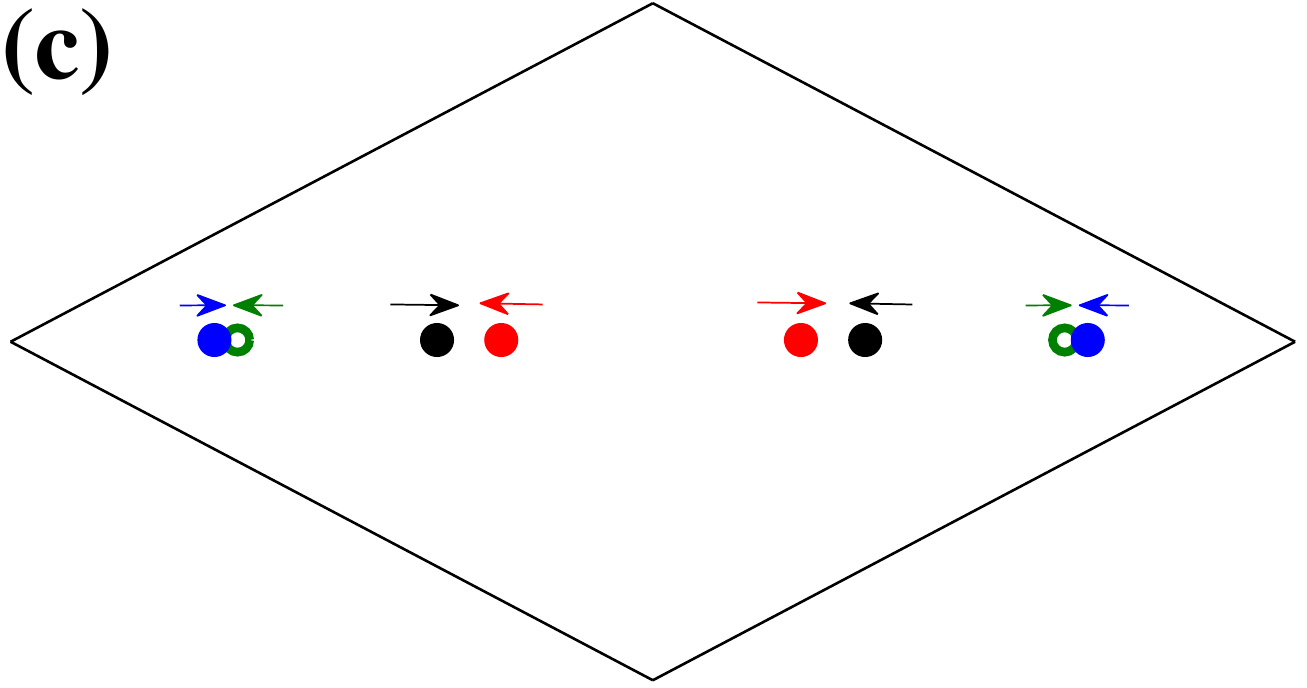}
\includegraphics[width=4.2cm]{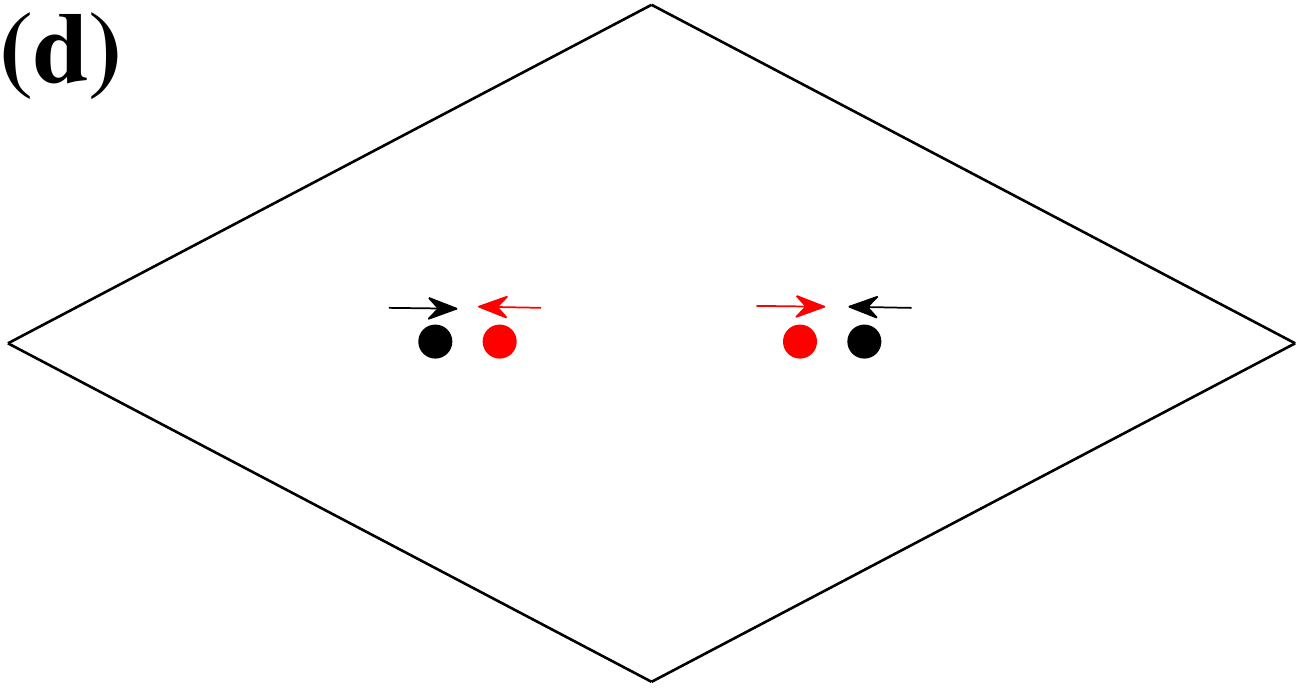}
\caption{Location of the cones in the field-induced gapless QSLs with $\pmb B\parallel(\hat x- \hat y)$.
Each dot (triangle) stands for a Majorana cone. The markers with the same color are symmetry related. The solid (hollow) dots or triangles stand for positive (negative) chirality. The arrows illustrate how the cones move with the increase of the field strength. When a pair of cones meet they merge and disappear, accompanied by a continuous phase transition [see Fig.~\ref{Bxy}(d)]. (a) The parent 20-cone PKSL state at $|\pmb B|=0$; the triangles stand for the cones that are gapped out as $|\pmb B|\neq0$.  (b) The field-induced 8-cone state at ${g\mu_B B/|K|}=0.42$.  (c) The field-induced 8-cone state at $ {g\mu_B B/|K|}=0.57$. (d) The field-induced 4-cone state at ${g\mu_B B/|K|}=0.61$; with further increasing of $B$ all of the cones will disappear and a gapped trivial phase will be obtained.
}
\label{BCones}{}{}
\end{figure}

Now we focus on the response of the ordered phase to magnetic fields. We only consider the FM and the zigzag order since the region of the size of the IS phase is very small. For the out-of-plane field with $\pmb B\parallel (\hat x+ \hat y+ \hat z)$, both the FM and the zigzag orders are suppressed by the field via first-order phase transitions after which the system enters the trivial phase [see Figs.~\ref{Bc} (c) and \ref{Bc}(d)]. Notice that the critical field of the zigzag phase in Fig.~\ref{Bc}(d) is large, and no intermediate CSL phase is found (we indeed obtain an intermediate state whose mean-field Chern number is $\nu=1$, but this state becomes trivial after Gutzwiller projection because its ground state degeneracy on a torus is 1 which indicates Z$_2$ confinement (see Appendix \ref{GSD} for details).

Then we apply in-plane magnetic field $\pmb B\parallel(\hat x-\hat y)$ to the ordered phases. Again, the FM phase and the trivial polarized phase are separated with a direct first-order transition [see Fig.~\ref{Bxy}(c)]. In contrast, in some region of the zigzag phase, after the magnetic order being suppressed by the field, an 8-cone phase and a 4-cone phase emerge in sequence with increasing $|\pmb B|$, as shown in Figs.~\ref{Bxy}(d) and \ref{BCones}(b)$\sim$\ref{BCones}(d).  The phase transition from the 8-cone phase to the 4-cone phase and the transition from the 4-cone phase to the trivial phase are both of second order, which are characterized by smooth changing of the variational parameters and the pairwise merging and disappearance of the Majorana cones [see Fig.~\ref{BCones}(c)].

To understand the nature of the field-induced gapless QSLs, we restore the $D_{3d}\times Z_2^T$ symmetry by removing the magnetic field manually while keeping all the other variational parameters intact, we find that the field-induced 8-cone state as well as the 4-cone state becomes a 20-cone PKSL (PKSL20) state. In other words, the two field-induced QSLs in Fig.~\ref{Bxy}(d) are descending from a PKSL20 phase.   Actually, at zero field the PKSL20 state is competing in energy with the zigzag state for $\Gamma'<0$ (the energy difference is of order $10^{-3}|K|\sim10^{-2}|K|$ per site). This helps to understand that a proper magnetic field can switch the ground state from the zigzag state to the descendants of the PKSL20 state, {\it i.e.}, the 8-cone state or the 4-cone state.
We note that no gapless QSLs are induced from the zigzag phase as $\Gamma$ term is small. Therefore, both $\Gamma$ and $\Gamma'$ interactions are important for the appearance and the robustness of the intermediate field-induced gapless QSLs.

\section{Discussions and Conclusions}\label{conclusion}
\subsection{Families of multinode Z$_2$ QSLs}\label{family}


\subsubsection{The ring-exchange interaction}

As shown in Fig.\ref{PSGs}, two PKSL8 states having the same PSG as the KSL and containing eight cones are competing in energy with the $\pi$-flux state.  The two PKSL8 states are labeled as PKSL8$_2$ and PKSL8$_4$ respectively, where the subscripts $_4$ and $_2$ stand for the Chern number of the resultant gapped states in a weak field along the $\hat c$-direction owing to their different chirality distributions of the cones. The difference between the two PKSL8 states can also be seen from their variational parameters listed in Table.~\ref{tab:PKSL8s}.

\begin{table}[htbp]
\centering
\begin{tabular}{c|ccccccc}
\hline
\hline
States  &  $\rho_a$  & $\rho_c$ & $\rho_d$ & $\rho_f$ & $\eta_0$ & $\eta_3$ & $\eta_5$  \\
\hline
PKSL8$_2$  & 0.1955  & -0.6929  & 0.4442  & -0.0108  & 0.2018  & 1.0188  & 0.2936 \\
PKSL8$_4$ & -0.0222  & -0.4958  & 0.3294  & -0.0044  & 0.1755  & 1.0323  & 0.2802 \\
\hline
\hline
\end{tabular}
\caption{Variational parameters of the PKSL8$_2$ and the PKSL8$_4$ states at fixed $\Gamma/|K|=0.6$ and $\Gamma'/|K|=0.2$.
}\label{tab:PKSL8s}
\end{table}

\begin{figure}[htbp]
\includegraphics[width=8.3cm]{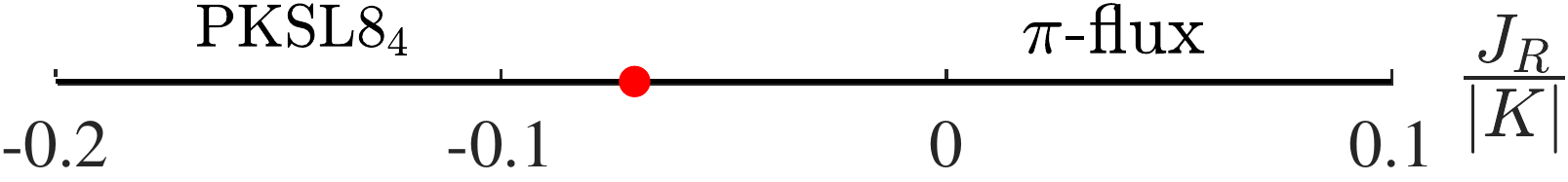}
\caption{Phase diagram of the $K$-$\Gamma$-$\Gamma'$-$J_{\rm R}$ model at fixed $\Gamma/|K|=0.6$ and $\Gamma'/|K| = 0.2$.
There is a direct first-order phase transition between $\pi$-flux state and PKSL8$_4$.}
\label{Ring}
\end{figure}

The competing PKSL8 states can be possibly stabilized by other interactions. To this end, we add the 6-spin ring-exchange interactions
$$H_{\rm r}=J_{\rm R}\sum_{\rm hexagon}\hat P_{\rm hexagon}$$
to the spin Hamiltonian, where
\[ \hat P_{\rm hexagon}=  -  \hat\chi_{ij}\hat\chi_{jk}\hat\chi_{kl}\hat\chi_{lm}\hat\chi_{mn}\hat\chi_{ni} - {\rm cyclic}(ijklmn) +  {\rm H.c.} \]
and $$\hat\chi_{ij}=C_i^\dagger C_j=c_{i\uparrow}^\dagger c_{j\uparrow} + c_{i\downarrow}^\dagger c_{j\downarrow},$$ with the index $i,j,k,l,m,n \in {\rm hexagon}$.

Therefore, the ring-exchange interactions can be easily decoupled using the parameter $\langle \hat\chi_{ij} \rangle$ which is already included in the matrix $U_{ji}^{(0)}$ (see Sec.\ref{SM:MFs}).
It turns out that one of the competing PKSL8 states is indeed stabilized for $J_{\rm R}/|K|\leq -0.07$ with $\Gamma/|K|=0.6$ and $\Gamma'/|K|=0.2$, as shown in Fig.~\ref{Ring}.

\subsubsection{Two families of multinode Z$_2$ QSLs}\label{sec:conenumber}
Above we show that for the Kitaev PSG there exist more than one gapless QSL with different numbers of cones in their excitation spectrum. Actually, the number of cones is restricted by symmetry. Notice that a general momentum point $\pmb k$ is invariant under the little co-group $\{E,PT\}$ whose coset contains $12/2=6$ elements. Generally, the representation in each coset transforms a cone into a new one. Therefore,  the cones locating at general $\pmb k$ points appear in multiples of 6.  The $\pmb K$ and $\pmb K'$ points are special since they are invariant under the little co-group $D_{3}\times \{E,PT\}$ and are transformed into each other by time-reversal $T$. Therefore, if there is a cone at $\pmb K$ (or $\pmb K'$) there must be a pair of them. Finally, the zone center $\pmb k=0$ which respects the fully $D_{3d}\times Z_2^T$ symmetry generally does not support a cone. Hence, a general gapless spin liquid contains $6n+2$ Majorana cones, where $n\geq0$ is an integer. The QSLs with different $n$ have different physical properties and can be distinguished by applying magnetic fields.  In the present work we realized three of them for PKSLs, with $n=0,1,2$, respectively.

A similar analysis can be applied to the $\pi$-flux states (see Appendix \ref{app:cones} for details). Noticing that the original BZ contains four copies of the compact BZs with the same dispersion, the number of Majorana cones in the family of $\pi$-flux QSL phases should be $4(6n+2), n\in\mathbb Z$. The one appearing in the phase diagram in Fig.~\ref{KGammaGammaPrime} has $n=1$.

\subsection{Conclusions}

We have studied the quantum $K$-$\Gamma$-$\Gamma'$ model on the honeycomb lattice using the variational Monte Carlo method. We find that as $K<0$ the non-Kitaev interactions $\Gamma>0$, $\Gamma'>0$ give rise to two gapless QSLs, one proximate Kitaev spin liquid phase which contains 14 Majorana cones (PKSL14), and $\pi$-flux states with 32 Majorana cones.  Although a similar $\pi$-flux state has been studied in literature\cite{fluxcrystal,SSZhang}, our work realized it in a simple lattice model which can be possibly realized in materials.  As the magnetic field $\pmb B\parallel (\hat x+ \hat y+ \hat z)$ is added to the gapless QSLs, three gapped CSLs (with Chern number $\nu=5, -8, 4$) are realized. For an applied in-plane weak magnetic field with $\pmb B\parallel (\hat x-\hat y)$,  a 6-cone (16-cone) field-induced gapless QSL phase is obtained from the PKSL14 ($\pi$-flux) phase. Interestingly, when the zigzag order is suppressed by the field $\pmb B\parallel (\hat x-\hat y)$, a gapless Z$_2$ QSL phase with 8 or 4 Majorana cones can be induced at intermediate field strength. This provides an alternative interpretation of the nuclear magnetic resonance experiment of $\alpha$-RuCl$_3$ \cite{rzea,Liu_KG}.

Interestingly, with the $K$, $\Gamma$, $\Gamma'$ and the negative six-spin ring-exchange interactions, we realized the GKSL, PKSL8, PKSL14 phases, which have the same PSG and belong to a big family of multinode Z$_2$ QSLs whose spinon excitation spectra contain $6n+2$ Majorana cones.  Furthermore, other families of nodal Z$_2$ QSLs with different PSGs also exist, for instance, the uniform $\pi$-flux QSLs contain $4(6n+2)$ 
cones. We trust that our study opens a door to seek different QSL phases in quantum magnets with strong spin-orbit couplings, and especially sheds light on the realization of various QSLs in experiments.

Finally, we mention a potential issue of our present work. Since we only considered a limited number of ansatz in our VMC approach, the variational states which we did not study may also be important. Especially, the ansatz with other PSGs deserve future exploration.

\section*{Acknowledgements}

We thank B. Normand, W. Li, H. Li and S.-S. Zhang for valuable discussions and comments. J.W., Q.Z., and Z.-X.L. are supported by the Ministry of Science and Technology of China (Grant No. 2016YFA0300504), the NSF of China (Grants No. 11574392 and No. 11974421), and the Fundamental Research Funds for the Central Universities and the Research Funds of Renmin University of China (No. 19XNLG11). X.W. is supported by MOST:  2016YFA0300501 and NSFC: 11974244 and additionally from a Shanghai talent program.

\appendix

\begin{widetext}
\section{Next-nearest-neighbor couplings}\label{NNNC}

In this section, we investigated the next-nearest-neighbor couplings in the mean-field Hamiltonian. Owing to the considered Kitaev PSG, the form of the next-nearest-neighbor couplings is significantly constrained. The most general expression of the mean-field Hamiltonian ansatz  with next-nearest-neighbor couplings is
\Beq
H_{\rm mf}^{\rm SL}  =  \sum_{\langle\langle i,j \rangle\rangle \in{\tilde\alpha}{\tilde\beta}({\tilde\gamma})} \!\!\! {\rm Tr}
\, [V_{ji}^{(0)} \psi_i^\dag \psi_j] \! + \! {\rm Tr} \, [V_{ji}^{(1)} \psi_i^\dag
(i R_{{\tilde\alpha}{\tilde\beta}}^{\tilde\gamma}) \psi_j]
 +  {\rm Tr} \, [V_{ji}^{(2)}
\psi_i^\dag \sigma^{\tilde\gamma} \psi_j] \! + \! {\rm Tr} \, [V_{ji}^{(3)} \psi_i^\dag
\sigma^{\tilde\gamma} R_{{\tilde\alpha}{\tilde\beta}}^{\tilde\gamma} \psi_j] \! + \! {\rm H.c.}
\Eeq
where $\langle\langle i,j \rangle\rangle$ denotes next-nearest-neighbor sites, ${\tilde\gamma}$ is used to label the type of the bond $\langle\langle i,j \rangle\rangle$ on the next-neighbor sites.  The Lagrangian multipliers are ignored again.

\begin{table*}[htbp]
\centering
\begin{tabular}{c|ccccccccccc}
\hline
\hline
 Energy         & $\rho_a$  & $\rho_c$ & $\rho_d$ & $\rho_f$ & $\eta_0$ & $\eta_3$ & $\eta_5$  & $\beta_0$ & $\beta_2$ & $\beta_4$ & $\beta_6$ \\
\hline
-0.61452496   & 0.6871 & -0.7244  & 0.6855  &  0.0034 & -0.0210 & -0.7707 & -0.0243  &   &   &   &  \\
\hline
-0.61446920   & 0.6867 & -0.7257  & 0.6871 &  0.0034 &  -0.0216 & -0.7741 &  -0.0349  &  0.0019 &  0.0020 &  0.0020  & 0.0022 \\
\hline
\hline
\end{tabular}
\caption{Variational parameters of the PKSL14 states at fixed $\Gamma/|K|=1.4$ and $\Gamma'/|K|=0.05$.
}\label{tab:PKSL14}
\end{table*}

Here we only consider the mean-field Hamiltonian based on the Kitaev PSG. The most general coefficients on next-nearest-neighbor sites preserving the $C_3$ rotation symmetry (in the PSG sense) also contain multiples of the uniform ($I$) and $\tau^x +\tau^y + \tau^z$ gauge components,
\Beq
&& V_{ji}^{(0)} = \beta_0 + i\beta_1 (\tau^x + \tau^y + \tau^z), \\
&& V_{ji}^{(1)} = i\beta_2 +  \beta_3 (\tau^x + \tau^y + \tau^z), \\
&& V_{ji}^{(2)} = i\beta_4 +  \beta_5 (\tau^x + \tau^y + \tau^z), \\
&& V_{ji}^{(3)} = i\beta_6 +  \beta_7 (\tau^x + \tau^y + \tau^z).
\Eeq
\end{widetext}
Thus a spin-liquid ansatz that preserves the full PSG symmetry generated by Eq.~(\ref{PSGKSL}) contains the variables $U_{ji}^{(m)} + V_{ji}^{(m)}$ with eleven real parameters $\rho_a$, $\rho_c$, $\rho_d$, $\rho_f$, $\eta_0$, $\eta_3$, $\eta_5$, $\beta_0$, $\beta_2$, $\beta_4$, $\beta_6$ allowed.
However, the next-neighbor-neighbor parameters ($\beta_0$, $\beta_2$, $\beta_4$, $\beta_6$) are vanishingly small from our VMC calculations (shown in Table.~\ref{tab:PKSL14}).

Considering that longer-range mean-field couplings are of even less significance compared to the next-nearest-neighbor terms,  we ignored them in our VMC calculations.\\

\section{First-order VS. continuous phase transitions}\label{transitions}

\subsection{Zero magnetic field}

The phase transition between PKSL14 and $\pi$-flux state in the phase diagram of the $K$-$\Gamma$-$\Gamma'$ model is of first order due to belonging different PSGs.
The numerical results also turn out to be a weak first order, which can be seen from the level crossing in the ground-state energy.

The phase transitions between the QSLs (namely the GKSL, the PKSL14, and the $\pi$-flux state ) and the magnetically ordered phases are sharply first-order, which are characterized by a sudden change of the variational order parameter $M$.  For example, at $\Gamma/|K|=1$ and $\Gamma'/|K|=0.04$ (the PKSL14 phase), we obtain $M\approx 8.9\times 10^{-3}$ which is nearly zero, but at $\Gamma/|K|=1$ and $\Gamma'/|K|=0.02$ (the zigzag phase), we get $M \approx 0.3932$, which is a finite number.

The phase transitions between magnetically ordered phases must be of first order because continuous phase transitions between different symmetry breaking orders are forbidden in the Landau paradigm.

\subsection{In a magnetic field $\pmb B\parallel (\hat x+\hat y+\hat z)$}

In this case, the time reversal symmetry ($T$) is broken by the field and the remaining symmetry group\cite{LuYuanMing} is $ \left(\mathscr C_3\rtimes \{E, TC_2\}\right) \times Z_2^P$, where $C_2$ is a twofold rotation whose axis lies in the lattice plane and $Z_2^P=\{E,P\}$ is the spatial inversion group.
The gapless QSLs are fully gapped out and become CSLs. In the following we only consider the transitions between CSLs.

If a phase transition between two CSLs is a continuous one, the spinon spectrum at the critical point must close its gap and form cone-like dispersions. Since no symmetry is broken at the transition point, the number of "cones" are restricted by symmetry. In the following we consider three possibilities:

(A) There is only one "cone" locating at $\pmb k=0$. Since one cone contributes either $1/2$ or $-1/2$ to the Chern number when it is gapped out, the Chern numbers at the two sides of the critical point differ by either 1 or $-1$.

(B) There are two "cones" locating at $\pmb K$ and $\pmb K'$. Since these two cones are related by inversion symmetry, they have the same contribution to the total Chern number. Therefore, at the transition point the Chern number may change by either $2$ or $-2$.

(C) There are six symmetry-related "cones" at general momentum points. At the transition point, the Chern number may change by either $6$ or $-6$.

\begin{figure}[t]
\includegraphics[width=7cm]{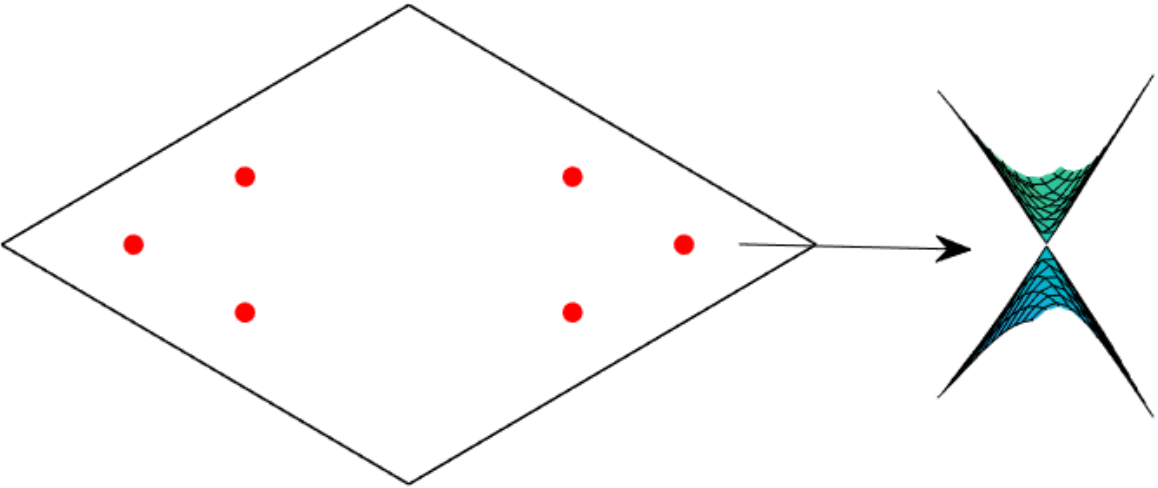}
\caption{
Spinon dispersion at the critical point between the $\nu=-8$ CSL and the $\nu=4$ CSL in Fig.~\ref{Bc}(b), where the six red dots stand for the gapless points in the compact BZ.
}
\label{transition}
\end{figure}

The transition from the $\nu=-8$ CSL to the $\nu=4$ CSL in Fig.~\ref{Bc}(b) is consistent with case (C) due to the fact that the magnetic BZ is equivalent to two identical copies of the compact BZ (see Appendix \ref{BZ} for details) and is indeed a continuous transition. The spinon dispersion at the critical point is illustrated in Fig.~\ref{transition}. \\

\begin{table}[b]
\centering
\begin{tabular}{c|ccccccc}
\hline
\hline
States  &  $\rho_a$  & $\rho_c$ & $\rho_d$ & $\rho_f$ & $\eta_0$ & $\eta_3$ & $\eta_5$  \\
\hline
6-cone  & 0.5325     &-0.5357   & 0.5171   & 0.0107   &-0.0499   &-0.5653   &-0.0205 \\
Trivial & -0.2090    &-0.5858   & 0.3832   & 0.0046   & 0.4935   & 0.2072   & 0.0771 \\
\hline
\hline
\end{tabular}
\caption{Variational paremters of the 6-cone state (for ${g\mu_B B/|K|}=0.07$) and the trivial polarized state (for ${g\mu_B B/|K|}=0.14$) in Fig.~\ref{Bxy}(a).
}\label{tab:Bxyparameters}
\end{table}

\subsection{In a magnetic field $\pmb B\parallel (\hat x-\hat y)$}
In this case, $C_3$ and $T$  are broken and the symmetry group reduces to $\{E, C_{2}T\}\times Z_2^P$, where the $C_2$ axis is perpendicular to $\pmb B$. Some of the cones in the original gapless QSLs are gapped out by the magnetic field and the rest are locating on the horizontal line, $i.e.$, the symmetric line of  $C_2T$. As stated in the main text, continuous transitions are characterized by pairwise merging and disappearance of the cones. We only list the first-order transitions below.

In Fig.~\ref{Bxy}(a), the transition from the 6-cone state to the trivial phase is first-order, where the variational parameters $\rho_a$, $\eta_0$, $\eta_3$ and $\eta_5$ reverse their sign (shown in Table.~\ref{tab:Bxyparameters}). The transitions from the FM to the trivial phase in Fig.~\ref{Bxy}(c) and from the zigzag phase to the 8-cone state in Fig.~\ref{Bxy}(d) are first-order, since the order parameter $M$ has a jump (not shown).\\

\section{Properties of the $\pi$-flux phase} \label{BZ}

\subsection{The magnetic BZ and the compact BZ}\label{app:cBZ}
\begin{figure}[t]
\includegraphics[width=4.05cm]{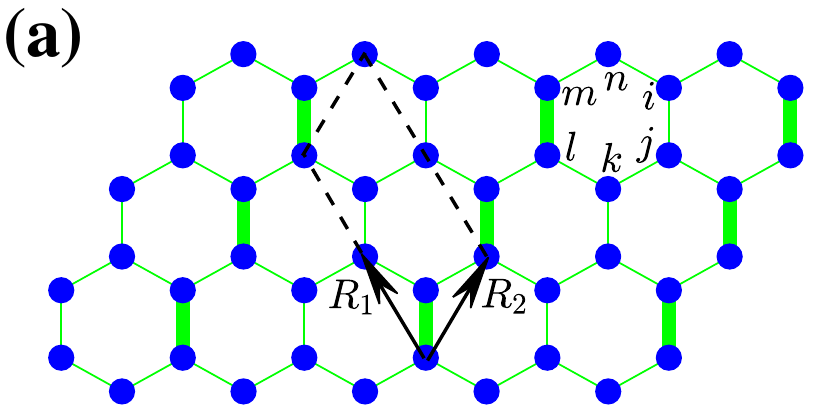} \
\includegraphics[width=4.2cm]{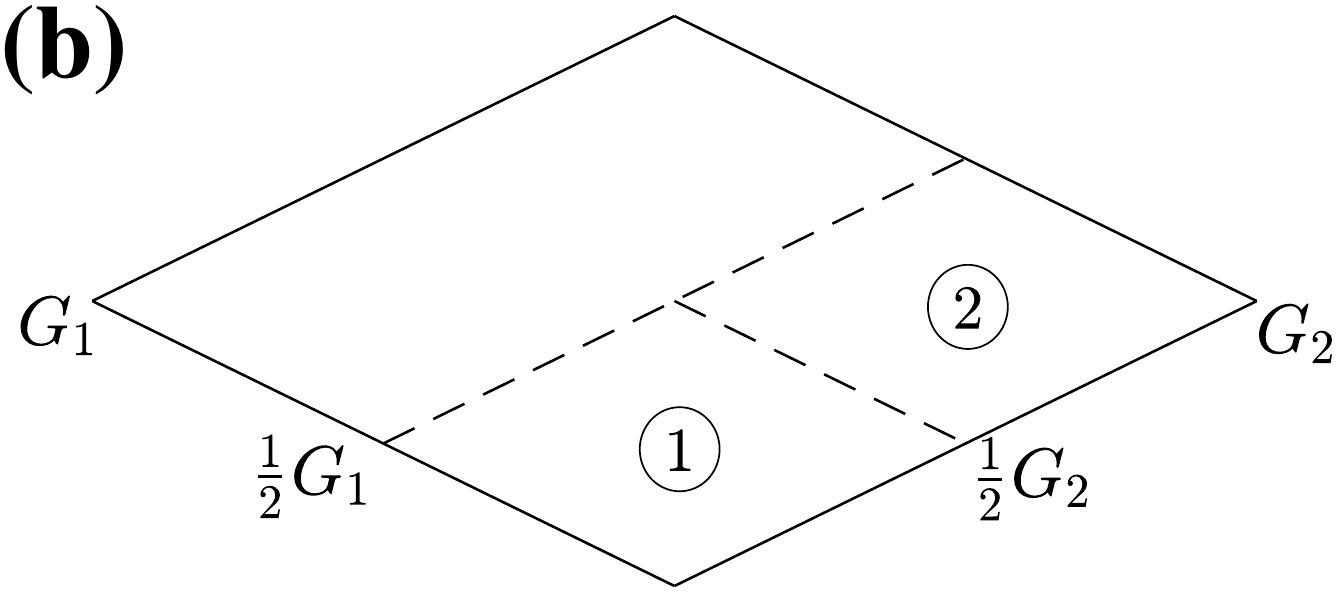}
\caption{(a) Configuration of the uniform $\pi$ flux on the honeycomb lattice. The rough line indicates an inverse of sign in the mean field. The original unit cell is spanned by the lattice vectors $\pmb R_{1,2}$ while the unit cell of the $\pi$ flux is marked by a black dashed parallelogram.
(b) The conventional BZ of zero-flux state is spanned by $\pmb G_1$ and $\pmb G_2$. The magnetic BZ of $\pi$-flux state is spanned by $\frac{1}{2}\pmb G_1$ and $\pmb G_2$ (the areas labeled as \ding{172} and \ding{173}), and the compact BZ is spanned by $\frac{1}{2}\pmb G_1$ and $\frac{1}{2}\pmb G_2$ (the area labeled as \ding{172}).
}
\label{lattice}
\end{figure}

In the mean-field Hamiltonian $H_{\pi\rm -flux}$ of the $\pi$-flux state, the unit cell is doubled [see Fig.~\ref{lattice}(a)]. The translation operation $\hat T_2$, which translates the sites by lattice vector $\pmb R_2$, is still a symmetry, but translation operation $\hat T_1$ which translates the sites by lattice vector $\pmb R_1$ is no longer a symmetry. Obviously $\hat T_1^2$ is a symmetry, which enlarges the unit cell and shrinks the original BZ to the magnetic BZ [see Fig.~\ref{lattice}(b)]. As a consequence, the period of the dispersion of $H_{\pi\rm -flux}$ along the $\pmb G_1$ direction is $\pmb G_1/2$.

Actually, the mean-field Hamiltonian is invariant under PSG operation $\hat T_1\hat g_{ T_1}$,
\[
\hat T_1\hat g_{ T_1} H_{\pi\rm -flux}(\hat T_1\hat g_{ T_1})^{-1} =H_{\pi\rm -flux}.
\]
Here $\hat g_{ T_1}$ is a gauge transformation sending the fermion $C_i$ into
\Beq
\hat g_{ T_1} C_i \hat g_{ T_1}^{-1} = (-1)^{p+q}C_i,
\Eeq
where the position vector $\pmb x_i$ of site $i$ is expanded as $\pmb x_i=p \pmb R_1+q\pmb R_2$. Equivalently, $\hat g_{ T_1}$ shifts the momentum of the fermions,
\Beq
\hat g_{ T_1} C_{\pmb k} \hat g_{ T_1}^{-1} = C_{\pmb k+{\pmb G_1/2}+{\pmb G_2/2}}.
\Eeq
Therefore, the gauge transformation $\hat g_{ T_1}$ exchanges the dispersions in the area \ding{172} and the area \ding{173} of the magnetic BZ (here we have used the $\pmb G_1/2$ and $\pmb G_2$ periods of the dispersion). Thus the obtained dispersion is the spectrum of
\[
\hat g_{ T_1} H_{\pi\rm -flux} \hat g_{ T_1}^{-1} = \hat T_1^{-1} H_{\pi\rm -flux} \hat T_1 =H'_{\pi\rm -flux}.
\]
$H_{\pi\rm -flux}$ and $H'_{\pi\rm -flux}$ are related by a global translation $\hat T_1$ which does not affect the momentum and energy (but does affect the wave function), hence they have the same dispersion. Therefore, we conclude that the dispersion of $H_{\pi\rm -flux}$ in area $\ding{172}$ is the same as that in area $\ding{173}$, namely, the dispersion has a twofold periodic structure along both $\pmb G_1$ and $\pmb G_2$ directions. For this reason, it is sufficient to illustrate the dispersion in the compact BZ [see Figs.~\ref{Cones}(b) and \ref{lattice}(b)].

%
%
%
%
%

In a magnetic field $\pmb B\parallel \hat c$, the $\pi$-flux state is turned into a CSL (supposing that the Z$_2$ gauge field is deconfined) whose mean-field Hamiltonian still contains a $\pi$-flux in each hexagon. The physical properties of the resultant CSL are determined by the mean-field Chern number $\nu$. For instance, the thermal Hall conductance $\kappa_{xy} = {\nu\over2}(\pi k_B^2 T/ 6h)$ is proportional to $\nu$. A subtle question is, should the Chern number $\nu$ be calculated from the compact BZ, the magnetic BZ, or the original BZ? Recalling that $\nu$ is a topological invariant for the fermionic bands in the mean-field theory in which the minimal period in the reciprocal lattice is the magnetic BZ, therefore we should adopt the magnetic BZ to compute $\nu$.

\subsection{Number of cones in the family of $\pi$-flux phases}\label{app:cones}

Noticing that the $\pi$ fluxes are uniformly distributed in the hexagons of the $\pi$-flux state, after Gutzwiller projection the translation symmetry is restored. Therefore, we should go back to the original BZ to observe all physical quantities. On the other hand, the Gutzwiller projection does not qualitatively change the spinon dispersion. Therefore, the number of cones in the spinon excitations of the $\pi$-flux QSL, which can be reflected in the spectrum of neutron scattering experiments, should be counted in the original BZ which is four times as large as that in the compact BZ (or twice as large as that in the magnetic BZ). For example, the $\pi$-flux state appearing in the phase diagram contains eight Majorana cones in the compact BZ, therefore there are 32 cones in the original first BZ.

It should be mentioned that the gauge transformations of the PSG operations in the $\pi$-flux state are site dependent. Therefore, generally speaking the positions of the cones are dependent on specific gauge choice. This indicates that the position of the cones cannot be observed directly. What is actually observable is the spin dynamical structure factor, whose low-energy signal contains the momentums and energies of the intra-cone and inter-cone particle-hole excitations\cite{PKSL} which are gauge invariant quantities. A symmetry $g$ is preserved in the excitation spectrum if the weight as a function of momentum transfer for a given frequency (energy transfer) is symmetric under the operation $g$. In the following, we will use the above property and the symmetry criterion to count the number of cones in a general  $\pi$-flux QSL state.

Now we briefly illustrate the number of cones in the family of $\pi$-flux phases (preserving the PSG given in Sec. \ref{sec:piflux}). We label the dark hollow dots in the compact BZ shown in Fig.~\ref{Cones}(b) as $\pmb K_1$ and $\pmb K_1'$ respectively, where
\Beq
\pmb K_1=\pmb K-{1\over4}(\pmb G_1+\pmb G_2),\\
\pmb K_1'=\pmb K' -{1\over4} (\pmb G_1+\pmb G_2)\Eeq
and $\pmb K={2\over3}{\pmb G_1}+{1\over3}{\pmb G_2}, \pmb K'={1\over3}{\pmb G_1}+{2\over3}{\pmb G_2}$. We further label the points $\pmb K_1+ {1\over2}\pmb G_2$,  $\pmb K_1+{1\over2}\pmb G_1$,  $\pmb K_1+{1\over2}(\pmb G_1+\pmb G_2)$ in the other three copies of compact BZs as $\pmb K_2, \pmb K_3,\pmb K_4$, respectively. Similarly $\pmb K_2', \pmb K_3'$ and $\pmb K_4'$ are defined.
In a general $\pi$-flux QSL phase, at least eight Majorana cones are found at momentums $\pmb K_1,\pmb K_2,\pmb K_3,\pmb K_4$ and $\pmb K_1',\pmb K_2', \pmb K_3',\pmb K_4'$ in the original BZ\cite{fluxcrystal}.  The cones at $\pmb K_1,\pmb K_2,\pmb K_3,\pmb K_4$ have completely the same dispersion ({\it i.e.} the same shape), thus can be considered as "equivalent" cones. The dispersion in the latter four cones at $\pmb K_1',\pmb K_2',\pmb K_3',\pmb K_4'$ are also "equivalent" and are related to the former ones by inversion symmetry (see the discussion below). However, the positions of the cones are not invariant under the physical $C_3$ rotation; it seems that the $C_3$ symmetry is not preserved in the $\pi$-flux phase.

Remember that only the momentum differences between the cones are observable. It is obvious that the set of vectors
\Beq
&&\{\pm(\pmb K_2-\pmb K_1), \pm(\pmb K_3-\pmb K_1),\pm(\pmb K_4-\pmb K_1)\}\\
&=&\{\pm(\pmb K_2'-\pmb K_1'), \pm(\pmb K_3'-\pmb K_1'),\pm(\pmb K_4'-\pmb K_1')\}\\
&=&\{\pm{\pmb G_2\over2}, \pm{\pmb G_1\over2}, \pm{(\pmb G_1+\pmb G_2)\over2}\}
\Eeq
is invariant under the $C_3$ operation up to a reciprocal lattice vector. This seems to guarantee that the spectrum of the inter-cone excitations between "equivalent" cones preserves the $C_3$ symmetry. Actually, a further requirement is that the shape of the cones are $C_3$ symmetric. For instance, we consider the following three excitations,
\Beq
&&\gamma_{\pmb K_1+\pmb p}^\dag \gamma_{\pmb K_2+\pmb q},\\
&&\gamma_{\pmb K_1+\hat C_3\pmb p}^\dag \gamma_{\pmb K_3+\hat C_3\pmb q},\\
&&\gamma_{\pmb K_1+\hat C_3^2\pmb p}^\dag \gamma_{\pmb K_4-(\pmb G_1+\pmb G_2)+\hat C_3^2\pmb q},
\Eeq
where $\gamma_{\pmb k}$ stands for the eigen-Bogoliubov particle with momentum $\pmb k$(we have omitted the band index), $\pmb p, \hat C_3\pmb p, \hat C_3^2\pmb p$ are the relative momentums at cone $\pmb K_1$, and $\pmb q,\hat C_3 \pmb q, \hat C_3^2\pmb q$ are the relative momentums at other three "equivalent" cones $\pmb K_2,\pmb K_3,\pmb K_4$, respectively. The momentum transfers of the three excitations are related by $C_3$ rotation, so the $C_3$ symmetry of the spectrum requires that the particles $\gamma_{\pmb K_1+\pmb p}, \gamma_{\pmb K_1+\hat C_3\pmb p}, \gamma_{\pmb K_1+\hat C_3^2\pmb p}$ have the same energy. Namely, the shape of the cone at $\pmb K_1$should be symmetric under $C_3$ rotation. A similar requirement holds for all of the eight cones.

Therefore, inter-cone excitations between "equivalent" cones are indeed $C_3$ symmetric and it is safe to fold the original BZ to the compact BZ. In other words, it is sufficient to analyze the excitations within a compact BZ. As long as the inter-cone excitations within a single compact BZ are $C_3$ invariant, the $C_3$ symmetry will be guaranteed in the whole BZ. So now we only need  to consider $\pmb K_1$ and $\pmb K_1'$. Since the vectors $\pm(\pmb K_1-\pmb K_1')$ themselves are also invariant under $C_3$ operation up to a reciprocal lattice vector, we can conclude that the inter-cone excitations between all of the eight cones at $\pmb K_1,\pmb K_2,\pmb K_3,\pmb K_4, \pmb K_1',\pmb K_2', \pmb K_3',\pmb K_4'$ indeed preserve the $C_3$ symmetry.



Now we consider inversion symmetry. Noticing that $-\pmb K_1=\pmb K_4'-{(\pmb G_1+\pmb G_2)}$ and that $\pmb K_4'$ is "equivalent" to $\pmb K_1'$, the dispersions at $\pmb K_1$ and $\pmb K_1'$ are related to each other by spatial inversion. To see this, we consider the following two excitations,
\Beq
&&\gamma_{\pmb K_1+\pmb p}^\dag \gamma_{\pmb K_1' + \pmb q},\\
&&\gamma_{\pmb K_1'-\pmb p}^\dag \gamma_{\pmb K_1 -\pmb q},
\Eeq
Their momentum transfers are opposite to each other, if the energy transfers are equal, then the inversion symmetry is preserved. This requires that the energies of the particles $\gamma_{\pmb K_1+\pmb p}$ and $\gamma_{\pmb K_1'-\pmb p}$ are the same, namely, the shape of the cones at $\pmb K_1$ and $\pmb K_1'$ are inversion images of each other.  Therefore, we can artificially consider $\pmb K_1'$ as the inversion partner (or time reversal partner) of $\pmb K_1$ in the compact BZ. Namely, we can restrict our discussion within the compact BZ as if the dispersion and the inter-cone excitation spectrum is inversion symmetric. All the other symmetry operations in the symmetry group $D_{3d}\times Z_2^T$ can be analyzed within the compact BZ in a similar way.



For the above reasons, if there are more cones in the compact BZ, we can treat $\pmb K_1, \pmb K_1'$ as $C_3$ rotation centers (the reason is that the shape of the cones at these points are $C_3$ symmetric) and regard them as inversion partners of each other, just like the $\pmb K$ and $\pmb K'$ points in the zero-flux phase. Then it follows that the number of cones is $6n+2$ (see Sec. \ref{sec:conenumber}) in the compact BZ, and is $4(6n+2)$ in the original BZ.

Notice that the symmetry group only requires the shape of the cones at the $C_3$ centers (namely, $\pmb K_1,\pmb K_2,\pmb K_3,\pmb K_4, \pmb K_1', \pmb K_2', \pmb K_3', \pmb K_4'$) to be $C_3$ symmetric.\\

\section{Ground state degeneracy}\label{GSD}

If the invariant gauge group (IGG) is Z$_2$, as in the KSL, fermion-pairing terms cannot be removed by any SU(2) gauge transformation. The Z$_2$ gauge-flux excitations in this case are usually gapped, and can remain deconfined even if the matter field is gapped with zero Chern number.
The confinement or deconfinement of the Z$_2$ gauge field is reflected in the ground state degeneracy (GSD) of the Gutzwiller projected state when placed on a torus. If the state is Z$_2$ confined (deconfined), then inserting a global Z$_2$ $\pi$ flux in one of the holes results in the same (a different) state. 

\begin{table}[b]
\centering
\begin{tabular}{c|c||c|c|c|c|c}
\hline
\hline
$\frac{g\mu_B B}{|K|}$ & $\, \nu \,$ & $\rho_1$ & $\rho_2$ & $\rho_3$ & $\rho_4$ & GSD \\
\hline
0.09  & $-8$  & 0.1832  & 0.7741  & 0.8063  & 2.2364 & 4  \\
\hline
0.87  & $4$  & 0.3886  & 0.4035  & 1.0000   & 2.2079 & 4 \\
\hline
2.94 & $ 0$  & $\,$8.2$\times$10$^{-13}$ & $\,$2.5$\times$10$^{-12}$ & $\,$1.5$\times$10$^{-10}$  & 4.0000 & 1 \\
\hline
\hline
\end{tabular}
\caption{Eigenvalues of the density matrices of the CSL ground states induced
by a magnetic field $\pmb B\parallel (\hat x+\hat y+\hat z)$, computed in the $\pi$-flux state ($\Gamma/
|K| =1$, $\Gamma'/|K| = 0.3$) for a system of size 8$\times$8$\times$2. $\nu$ is the
mean-field Chern number.}
\label{tab:GSD}
\end{table}

Because this process is equivalent to exchanging the boundary conditions of the mean-field Hamiltonian from periodic to anti-periodic, in two dimensions one may construct the four mean-field ground states $|\psi_{\pm\pm} \rangle$, where the subscripts denote the boundary conditions for the $x$- and $y$-directions. After a Gutzwiller projection of these four states to the physical Hilbert space, the number of linearly independent states is equal to the GSD on a torus.

To make sure that field-induced CSLs are nontrivial, we calculate the density matrix of the projected (VMC) states from the wave-function overlap $\rho_{\alpha\beta} = \langle P_G \psi_\alpha | P_G \psi_\beta \rangle = \rho_{\beta\alpha}^*$, with $\alpha,\beta \in \{++,+-,-+,--\}$.
If $\rho$ has only one significant eigenvalue, with the others vanishing, then the GSD is 1, indicating that the Z$_2$ gauge field is confined. If $\rho$ has more than one near-degenerate nonzero eigenvalue, the GSD is nontrivial and hence the Z$_2$ gauge fluctuations are deconfined. In the deconfined phases, if the Chern number is even then from above the GSD is 4; however, if the Chern number is odd, then the GSD is 3 because the mean-field ground state $|\psi_{++} \rangle$ has odd fermionic parity and vanishes after Gutzwiller projection. The field-induced CSLs with $\nu=-8$ and $\nu=4$ shown in Fig.~\ref{Bc}(b) are deconfined, whose GSD information (for a system with $8\times 8$ unit cells) is shown in Table.~\ref{tab:GSD}.

\begin{figure}[t]
\includegraphics[width=8.5cm]{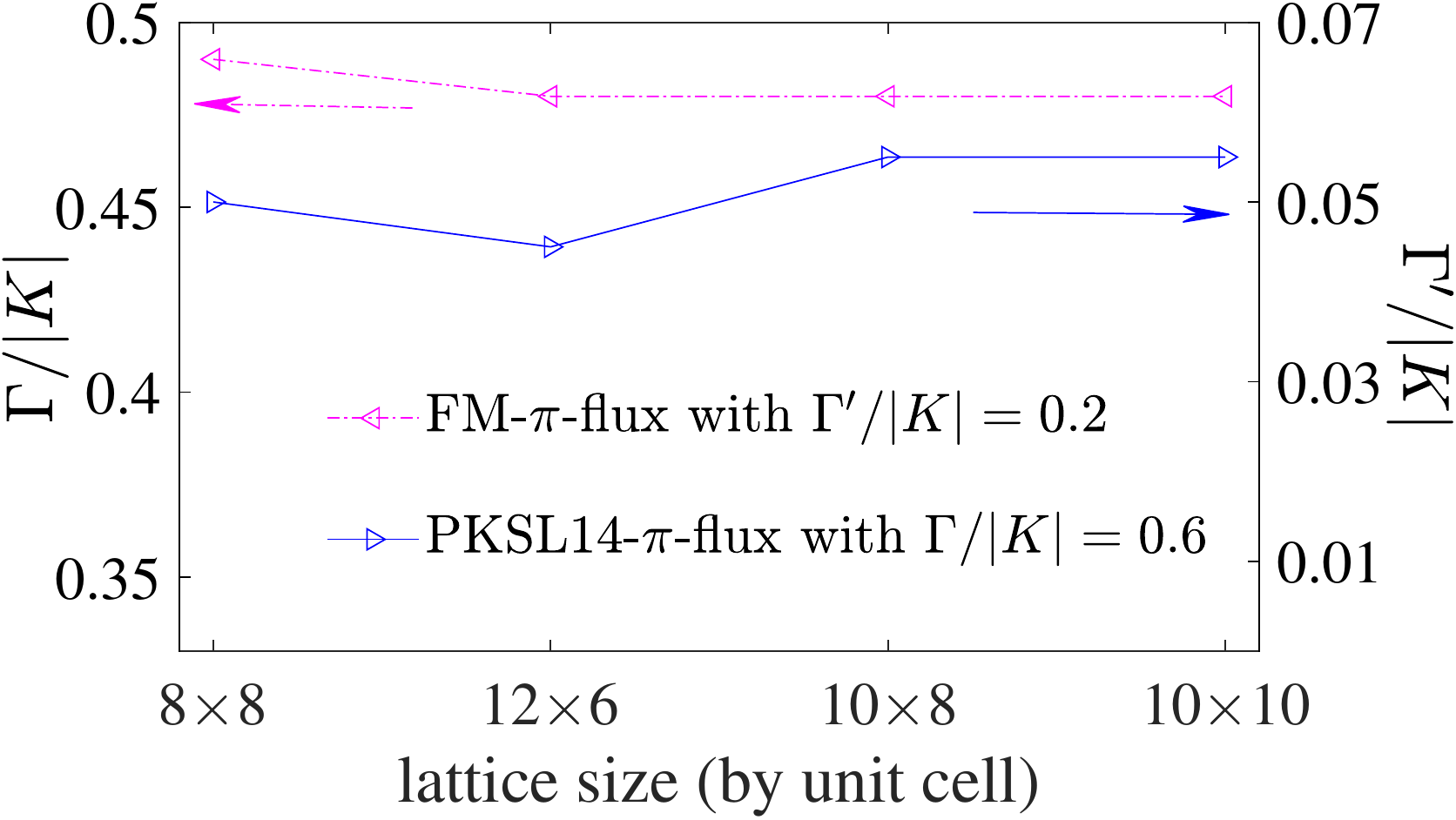}
\caption{Dependence on system size of the phase boundary between the FM ordered phase and the
$\pi$-flux phase, computed at fixed $\Gamma'/|K| = 0.2$, and of the boundary between the
PKSL14 phase and the  $\pi$-flux phase, computed at fixed $\Gamma/|K| = 0.6$.}
\label{boundary}
\end{figure}

\begin{figure}[b]
\includegraphics[width=8.1cm,]{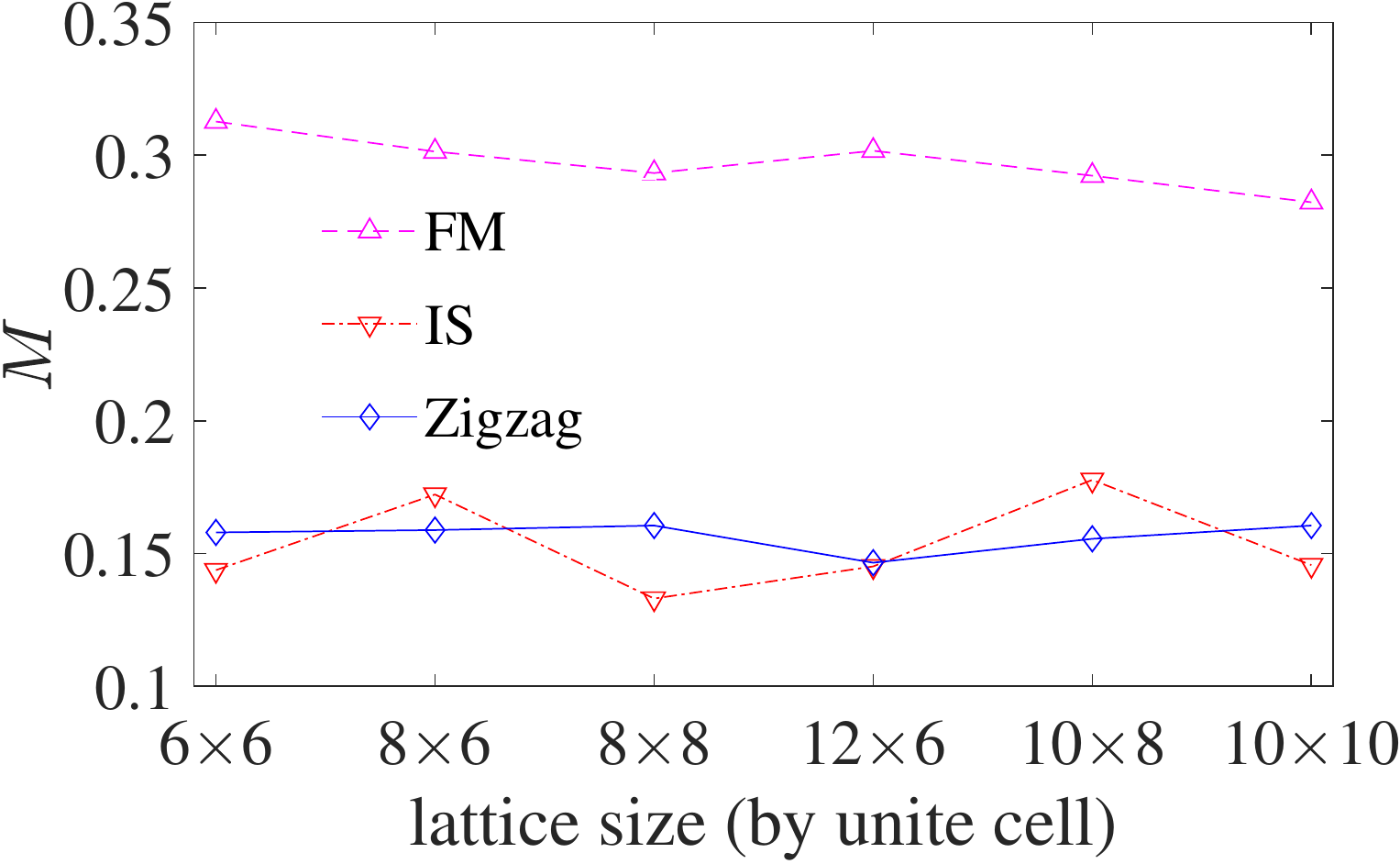}
\caption{Size scaling of the magnitude of $M$ for ordered states,
such as FM ($\Gamma/|K|=0.1$, $\Gamma'/|K|=0.2$), IS ($\Gamma/|K|=0.6$), and zigzag ($\Gamma/|K|=0.6$, $\Gamma'/|K|=-0.05$).
}
\label{orders}
\end{figure}

However, the field $\pmb B\parallel (\hat x+\hat y+\hat z)$ induced "intermediate state" (whose mean-field Chern number is $\nu=1$) based on the zigzag ordered state [shown in Fig.~\ref{Bc}(d)] becomes confined after Gutzwiller projection and belongs to the trivial polarized phase. The eigenvalues of the density matrices (for a system with $8\times 8$ unit cells) are given by 5.8$\times$10$^{-6}$, 7.0$\times$10$^{-4}$, and 2.9993, with only one non-vanishing value, indicating that the GSD on a torus is 1.

\section{Finite-size effect}\label{finitesize}

To see the finite-size effects of the VMC method, we calculated the size dependence of the phase boundaries. The phase boundaries from the $\pi$-flux phase to the PKSL14 phase and the ferromagnetic (FM) phase are shown in Fig.~\ref{boundary}. The slight size dependence indicates that the phase diagram is qualitatively the same in the thermodynamic limit.

We also perform a size-scaling of the magnitude of $M$ in Eq.~(\ref{Order}) in the three ordered phases,  as shown in Fig.~\ref{orders}. Although $M$ is slightly fluctuating with system size, they do not seem to vanish in the large size limit.




\begin{thebibliography}{99}

\bibitem{Balents}
L. Balents,
Nature (London) {\bf 464}, 199 (2010).

\bibitem{ZhouYi}
Y. Zhou, K. Kanoda, and T.-K. Ng,
Rev. Mod. Phys. {\bf 89}, 025003 (2017).

\bibitem{Kitaev}
A. Kitaev, Ann. Phys. {\bf 321}, 2 (2006).


\bibitem{rjk}
G. Jackeli and G. Khaliullin,
Phys. Rev. Lett. {\bf 102}, 017205 (2009).


\bibitem{rcjk}
J. Chaloupka, G. Jackeli, and G. Khaliullin,
Phys. Rev. Lett. {\bf 105}, 027204 (2010).

\bibitem{rsea}
J. A. Sears, M. Songvilay, K. W. Plumb, J. P. Clancy, Y. Qiu, Y. Zhao,
D. Parshall, and Y.-J. Kim,
Phys. Rev. B {\bf 91}, 144420 (2015).

\bibitem{rjea}
R. D. Johnson, S. C. Williams, A. A. Haghighirad, J. Singleton, V. Zapf, P.
Manuel, I. I. Mazin, Y. Li, H. O. Jeschke, R. Valent\'{\i}, and R. Coldea,
Phys. Rev. B {\bf 92}, 235119 (2015).

\bibitem{rcaoetal}
H.-B. Cao, A. Banerjee, J.-Q. Yan, C. A. Bridges, M. D. Lumsden, D. G.
Mandrus, D. A. Tennant, B. C. Chakoumakos, and S. E. Nagler,
Phys. Rev. B {\bf 93}, 134423 (2016).


\bibitem{ryea}
F. Ye, S.-X. Chi, H.-B. Cao, B. C. Chakoumakos, J. A. Fernandez-Baca,
R. Custelcean, T.-F. Qi, O. B. Korneta, and G. Cao,
Phys. Rev. B {\bf 85}, 180403(R) (2012).


\bibitem{rchoietal}
S. K. Choi, R. Coldea, A. N. Kolmogorov, T. Lancaster, I. I. Mazin, S. J.
Blundell, P. G. Radaelli, Y. Singh, P. Gegenwart, K. R. Choi, S.-W. Cheong,
P. J. Baker, C. Stock, and J. Taylor,
Phys. Rev. Lett. {\bf 108}, 127204 (2012).


\bibitem{Williams}
S. C. Williams, R. D. Johnson, F. Freund, S. Choi, A. Jesche, I. Kimchi, S. Manni, A. Bombardi, P. Manuel, P. Gegenwart, and R. Coldea,
Phys. Rev. B {\bf 93}, 195158 (2016).


\bibitem{lukas}
L. Janssen, E. C. Andrade, and M. Vojta,
Phys. Rev. B {\bf 96}, 064430 (2017).

\bibitem{Jinsheng}
K. Ran, J. Wang, W. Wang, Z.-Y. Dong, X. Ren, S. Bao, S. Li, Z. Ma, Y. Gan,
Y. Zhang, J. T. Park, G. Deng, S. Danilkin, S.-L. Yu, J.-X. Li, and J. Wen,
Phys. Rev. Lett. {\bf 118}, 107203 (2017).

\bibitem{npj}
P. Laurell and S. Okamoto,
npj Quantum Mater. {\bf 5}, 2 (2020).


\bibitem{KJG_zigzag}
S. M. Winter, K. Riedl, P. A. Maksimov, A. L. Chernyshev, A. Honecker, and R. Valent\'{\i},
Nat Commun {\bf 8}, 1152 (2017).

\bibitem{You_zig}
I. Kimchi and Y.-Z. You,
Phys. Rev. B {\bf 84}, 180407 (2011).

\bibitem{Thomale}
Y. Singh, S. Manni, J. Reuther, T. Berlijn, R. Thomale, W. Ku, S. Trebst, and P. Gegenwart,
Phys. Rev. Lett. {\bf 108}, 127203 (2012).



\bibitem{Valenti}
S. M. Winter, Y. Li, H. O. Jeschke, and R. Valent\'{\i},
Phys. Rev. B {\bf93}, 214431 (2016).

\bibitem{Rau}
J. G. Rau and H.-Y. Kee,
arXiv:1408.4811.

\bibitem{hidden}
J. Chaloupka and G. Khaliullin,
Phys. Rev. B {\bf 92}, 024413 (2015).

\bibitem{HYKee}
J. S. Gordon, A. Catuneanu, E. S. S\/{\o}rensen, and H.-Y. Kee,
Nat Commun {\bf 10}, 2470 (2019).

\bibitem{tensor}
H.-Y. Lee, R. Kaneko, L.-E. Chern, T. Okubo, Y. Yamaji, N. Kawashima, and Y.-B. Kim,
Nat Commun {\bf 11}, 1639 (2020).

\bibitem{igg}
X.-G. Wen, Phys. Rev. B {\bf 65}, 165113 (2002).

\bibitem{You_PSG}
Y.-Z. You, I. Kimchi, and A. Vishwanath,
Phys. Rev. B {\bf 86}, 085145 (2012).

\bibitem{PKSL}
J. Wang, B. Normand, and Z.-X. Liu,
Phys. Rev. Lett. {\bf 123}, 197201 (2019).

\bibitem{rins}
A. Banerjee, J. Q. Yan, J. Knolle, C. A. Bridges, M. B. Stone, M. D. Lumsden,
D. G. Mandrus, D. A. Tennant, R. Moessner, and S. E. Nagler,
Science {\bf 356}, 1055 (2017).

\bibitem{rwdyl}
W. Wang, Z.-Y. Dong, S.-L. Yu, and J.-X. Li,
Phys. Rev. B {\bf 96}, 115103 (2017).

\bibitem{rcm}
J. Cookmeyer and J. E. Moore,
Phys. Rev. B {\bf 98}, 060412(R) (2018).

\bibitem{Anderson88}
I. Affleck, Z. Zou, T. Hsu, and P. W. Anderson,
Phys. Rev. B {\bf 38}, 745 (1988).



\bibitem{Liu_KG}
Z.-X. Liu and B. Normand,
Phys. Rev. Lett. {\bf 120}, 187201 (2018).

\bibitem{Aniso}
J. Wang and Z.-X. Liu,
Phys. Rev. B {\bf 102}, 094416 (2020).


\bibitem{singleQ}
J. G. Rau, E. K.-H. Lee, and H.-Y. Kee,
Phys. Rev. Lett. {\bf 112}, 077204 (2014).



\bibitem{class}
L.-E. Chern, R. Kaneko, H.-Y. Lee, and Y.-B. Kim,
Phys. Rev. Research {\bf 2}, 013014 (2020).



\bibitem{fluxcrystal}
S.-S. Zhang, C. D. Batista, and G. B. Hal\'{a}sz,
Phys. Rev. Research {\bf 2}, 023334 (2020).









\bibitem{SSZhang}
S.-S. Zhang, Z. Wang, G. B. Hal\'{a}sz, and C. D. Batista,
Phys. Rev. Lett. {\bf 123}, 057201 (2019).


\bibitem{rzea}
J. Zheng, K. Ran, T. Li, J. Wang, P.-S. Wang, B. Liu, Z.-X. Liu, B. Normand,
J. Wen, and W. Yu,
Phys. Rev. Lett. {\bf 119}, 227208 (2017).





\bibitem{LuYuanMing}
H.-C. Jiang, C.-Y. Wang, B. Huang, and Y.-M. Lu,
arXiv:1809.08247.



\end{thebibliography}
\end{document}